\documentclass[11pt,draftclsnofoot,onecolumn]{IEEEtran}
\usepackage[colorlinks,
linkcolor=red,
anchorcolor=green,
citecolor=blue%green%blue
]{hyperref} % Add hyperlinks
\usepackage{amsmath,amssymb}
\usepackage{latexsym}
\usepackage{CJK}
\usepackage{cite}

\usepackage{color, soul}

\usepackage{graphicx}
\usepackage{epstopdf}
\usepackage{epsfig}
\usepackage{subfigure} % ͼƬ²¢ÅÅÏÔʾ

\usepackage{verbatim}  % \begin{comment} ... \end{comment}

\usepackage{mathrsfs}	
\usepackage{algorithm} %format of the algorithm
\usepackage{algorithmic} %format of the algorithm
\usepackage{booktabs}
\usepackage{textcomp}
\usepackage{multirow}
\usepackage{lettrine}

\usepackage{mathrsfs}
\usepackage{amsmath}
\usepackage{amssymb}

\usepackage{algorithm}
\usepackage{algorithmic}
\usepackage{extarrows}
\usepackage{colortbl}
\definecolor{mygray}{gray}{.9}
\usepackage{amsthm} %thm / proof

{ \bgroup
  \addtolength\abovedisplayshortskip{#1}
  \addtolength\abovedisplayskip{#1}
  \addtolength\belowdisplayshortskip{#1}
  \addtolength\belowdisplayskip{#1}}
{\egroup\ignorespacesafterend}

\usepackage{subeqnarray}%×Ó¹«Ê½½øÐбàºÅ
\usepackage{cases}%À¨ºÅÁªÁ¢ºê°ü
\usepackage{mathtools} %%Ëõ¶Ì¹«Ê½×Öĸ¼äµÄ¾àÀë
\usepackage{bm}%%¹«Ê½ÖмӴֺê°ü
\usepackage{color}
\definecolor{shadecolor}{rgb}{1,0.94,0}
\usepackage{framed}

\ifCLASSINFOpdf
\else
\fi
\begin{document}
\title{Importance of Small Probability Events in Big Data: Information Measures, Applications,\\ and Challenges}
\author{Rui~She,
        Shanyun~Liu,
        Shuo~Wan,
        Ke~Xiong,~\IEEEmembership{Member,~IEEE,}
        and~Pingyi~Fan,~\IEEEmembership{Senior~Member,~IEEE}% <-this % stops a space
%\thanks{
%This work is supported by the National Natural Science Foundation of China (NSFC) No. 61771283.}
\thanks{
R. She, S. Liu, S. Wan and P. Fan are with Beijing National Research Center for Information Science and Technology, and the Department of Electronic Engineering, Tsinghua University, Beijing, 100084, China (e-mail: sher15@mails.tsinghua.edu.cn (R.S.); liushany16@mails.tsinghua.edu.cn (S.L.); wan-s17@mails.tsinghua.edu.cn (S. W.); fpy@tsinghua.edu.cn (P.F.)).}
\thanks{
K. Xiong is with Beijing Key Laboratory of Traffic Data Analysis and Mining, and the School of Computer and Information Technology, Beijing Jiaotong University, Beijing 100044, China (e-mail: kxiong@bjtu.edu.cn).}
}
\maketitle

\begin{abstract}
In many applications (e.g., anomaly detection and security systems) of smart cities, rare events dominate the importance of the total information of big data collected by Internet of Things (IoTs).
That is, it is pretty crucial to explore the valuable information associated with the rare events involved in minority subsets of the voluminous amounts of data.
To do so, how to effectively measure the information with importance of the small probability events from the perspective of information theory is a fundamental question.
This paper first makes a survey of some theories and models with respect to importance measures and investigates
the relationship between subjective or semantic importance and rare events in big data.
Moreover, some applications for message processing and data analysis are discussed in the viewpoint of information measures.
{In addition, based on rare events detection, some open challenges related to information measures, such as smart cities, autonomous driving, and anomaly detection in IoTs, are introduced which can be considered as future research directions}.
\end{abstract}

\IEEEpeerreviewmaketitle

\label{sec:introduction}
{I}{t} is predicted that by 2050, the urban population all over the world is going to be doubled, which will rise up to 6.7 million people. With the rapid growth in the amount of urban residents, cities bring in new opportunities, while many new challenges come up, including environmental deterioration, sanitation problem, traffic congestion and terrorist attacks. In order to figure out these problems so that citizens may enjoy a new daily life with security and convenience, Internet of Things (IoTs) has been emerging as an effective solution \cite{ref_Big-IoT-data-analytics,ref_Big-Data-Deep-Learning,ref_Toward-scalable-systems-for-big-data-analytics,
ref_Machine-learning-with-big-data,ref_Information-security-in-big-data}.

{In IoTs, explosively increasing sensors and devices are deployed to sense and collect different types of data, e.g., states of moving cars, crossroads and subway tracks, which drive us into a ``big data'' era.}
{In order to make things smart, massive data has to be mined to find useful information and knowledge. In this case, the key point lies in how to deal with the observed data and dig out the hidden valuable information} \cite{Lu_Fan,X_Fan,X2_Fan,X3_Fan,ref_Deep-learning,ref_DNN-filter-bank,ref_Text-independent-speaker,
ref_fingerprinting-fusion}.
To do so,
a series of promising technologies have been put forward such as statistical learning, computer vision, signal processing and so on \cite{ref_Variational-bayesian-matrix,ref_Decorrelation-of-neutral,ref_Vector-quantization,
ref_Network-denoising,ref_Making-better-use,
ref_Data-intensive-applications,
ref_Preconditioned-data-sparsification}.

\subsection{Importance of Rare Events with Small Probability}\label{section1_0}

As a matter of fact, in some applications, the regular patterns of systems' or users' behaviours are required to be explored from common events that often occur, but for the other applications, the rare events attract more attention than those occurring with large probability.
For example, in financial crime detection systems, only a few illegal identities causing financial frauds indeed catch our eyes \cite{ref_A-comprehensive-survey-of-data-mining-based-accounting-fraud-detection-research,
ref_Fraudulent-Financial-Reporting},
which are more important from subjective consciousness.
Besides, in intrusion detection systems, only a few number of security alarms should be detected and handled
\cite{ref_Data-Mining-Approaches-for-Intrusion-Detection,
ref_Mining-intrusion-detection-alarms-for-actionable-knowledge,
ref_Research-on-intrusion-detection,
ref_Application-of-CART-decision,
ref_Neural-network-based}.

So far,
a lot of works have investigated networking intrusion and reliable communications to protect IoTs from being attacked
\cite{ref_Event-analysis-for-security-incident-management,
ref_Securing-embedded-systems,
ref_Security-in-embedded-systems,
ref_Blockchains-and-Smart-Contracts,
ref_Design-and-realization,
ref_A-Survey-on-Internet,
ref_Internet-of-things-and-big-data,
ref_SAlightweight-anomaly-detection-technique,
ref_Performance-and-accuracy-trade-off-analysis,
ref_SVELTE,
ref_A-lightweight-anomaly-detection-technique}, which show that the rare events should be focused on for their special value in IoTs.
By resorting to IoTs or other monitoring devices \cite{ref_Anomaly-detection-in-environmental},
smart city is becoming a timing fashion in city planning, construction, management and operations
%hot topic for many researches
\cite{ref_An-anomaly-detection-in-smart-cities,
ref_Internet-of-Things-for-Smart-Cities,
ref_Dynamic-network-model-for-smart-city,
ref_Flexible-spectrum-management-in-a-smart-city,
ref_Smart-city,
ref_To-Smart-City}.
%ieee access others
In this case, the rare events observed from monitoring systems also contain more significant features in the numerous data, which can provide effective references
% for the many applications including
for transportation management, city planning and public safety.

%\subsection{Modern researches for rare events in IoTs and  big data}
%%====================================================================================
%In practice, anomaly detection regarded as a kind of rare events detection is imperative for the security of IoTs \cite{Event-analysis-for-security-incident-management,
%Securing-embedded-systems,Security-in-embedded-systems}.
Due to the fact that anomalous events may be hidden in big data \cite{ref_Efficient-algorithms-for-mining-outliers-from-large-data-sets,
ref_Improved-principal-component,
ref_An-improved-methodology,
ref_Nonlinear-Gaussian-belief,
ref_Principal-components-selection}, it is significant to process rare events or the minorities in objective detection.
With regard to the autonomous driving in highways, it is crucial to detect the unexpected moving obstacles
over lanes (which can be viewed as rare events).
{It is reported that around 150 people die from road hazards in American traffic accidents every year}
\cite{ref_Traffic-safety-facts,ref_Detecting-unexpected-obstacles-for-self-driving-cars}.
%Lane Following and Obstacle Detection Techniques in Autonomous Driving Vehicles
{It is beneficial to develop autonomous driving cars based on anomalous objective detection in many aspects such as reducing traffic congestion and accidents, improving energy efficiency and ensuring transportation safety.}
Actually, there are some researches trying to design intelligent vehicle systems to avoid dangerous driving events with small probability \cite{ref_Lane-following-and-obstacle-detection-techniques,
ref_An-improved-lane-departure-method,
ref_Implementation-of-lane-detection-system,
ref_Region-based-convolutional-networks,
ref_Object-detection-with,
ref_Traffic-flow-prediction,
ref_Real-time-obstacles-detection-and-status-classification}.

In brief, rare events have special values in many newly rising fields such as IoTs,
smart city, and autonomous driving. Actually, the approaches for small probability processing are investigated from many perspectives in big data era.

\subsection{Information Theory for Rare Events}\label{section1_1}
In the viewpoint of information theory, information measures could have a seat on the table of rare events processing in big data.
According to conventional information theory, the uncertainty of probability distributions can be characterized by information measures such as Kolomogorov complexity, Shannon entropy, Renyi entropy, and mutual information. These measures are also applicable to the infrequent or abnormal events
\cite{ref_Anomaly-detection-A-survey,
ref_Multiblock-independent-component}.
%<<Anomaly Detection: A Survey>>
By using information measures to analyze the complexity of the different classes in big data, rare events would be recognized and handled \cite{ref_Information-theoretic-measures-for-anomaly-detection}.
For instance, an objective function of distribution was proposed based on factorization to detect the subsets with smaller probability
\cite{ref_An-information-theoretic-approach-to-detection-of-minority-subsets-in-database}.
Additionally, as an effective information distance, the relative (or differential) entropy is also applied to outlier detection
%MI divergence
%Yuheng Bu
\cite{ref_Kullback-Leibler-Divergence-(KLD)-Based-Anomaly-Detection}.
 %\cite{Nonparametric-divergence-estimation-with-applications}.
Although there are special scenarios where the above approaches can be used, it is evident that they just focus on the large probability elements or subsets to deal with rare events.

From the perspective of small probability elements,
there are also some technologies in the framework of statistical mechanics, such as the large deviation approaches and the measure of concentration of rare events
\cite{ref_The-large-deviation,ref_A-measure-of-the-concentration}.
In these cases, traditional information measures are explained and extended by aiming at minority subsets processing.
These technologies could be also used in many applications such as secure lossy compression and anomaly detection
\cite{ref_A-large-deviations-approach,
ref_statistical-traffic-anomaly-detection}.

In the framework of data distribution processing, the information divergence as a kind of information measure is an intersection of information theory and big data analytics.
In fact, information divergences can be adopted to measure the distance between two distributions with small probability elements.
Currently, information divergences have been used in many applications involved with rare events such as faulty detection \cite{ref_Analytical-model-of-the-KL-Divergence}, key frame selection \cite{ref_Key-frame-selection-based-on-KL-divergence} and
%text classification \cite{Feature-selection-algorithm-for-hierarchical-text-classification},
image recognition \cite{ref_Human-Ear-recognition-based-on-Multi-scale},
\cite{ref_A-Study-on-Invariance-of-$f$-Divergence-and-Its-Application}.
Therefore, how to use information measures to cope with small probability events becomes more interesting.

\subsection{The motivations and contributions}

The purpose of this paper is to integrate the works on importance analysis of small probability events and clarify the relationship between small probability cases with more importance and information processing including the corresponding information measures and applications. Essentially, this paper is not a technical work but a survey to summarize some classical theories and approaches of information processing based on small probability events so that the related literature can be discovered in a logical and reasonable way.

As far as the contribution of this work is concerned, a theoretical framework with a common fundamental form of message importance measure is constructed to show the core idea of importance of small probability events and characterize its mathematical representation.
Moreover, similar to Shannon entropy, an information processing architecture is proposed from the perspective of message importance to combine the message importance compression, transmission loss and receiver preprocessing, which may broaden the extension of conventional information theory.
In this case, some novel source coding strategies and information distortion analysis are obtained in an information system based on the importance of small probability events.
For big data analytics, some related technologies including measures estimation, dimension reduction and correlation analysis are also unified into an architecture of information system to process important small probability events. This provides a reasonable data processing procedure for the small probability events hidden in massive samples.
{Finally, some modern and challenging applications, such as smart cities, autonomous driving, and IoTs, may adopt the information measures based on the message importance as novel criterions or metrics for rare events detection. In this regard, we present some schemes with information measures for the corresponding applications.
}

\subsection{Organization}
The organization of the rest parts of this paper is summarized as follows.
In Section \ref{section2}, we analyze some theories and technologies of information measures in the scenarios where rare events have valuable sense.
%such as the message importance measure (MIM) with parameter selection and the non-parametric MIM.
In Section \ref{section3}, we discuss some applications based on information measures for rare events, including information compression, transmission and preprocessing.
Section \ref{section4} first introduce some effective estimations of distributions and their functionals.
Then, information coupling, directed information and some applications involved with rare events are introduced to reduce the dimension of big data and analyze the data causality or correlation from the perspective of information theory.
In Section \ref{section5}, {some challenging research directions for information measures are presented based on the rare events detection.}
At last, we conclude the paper in Section \ref{section6}.

\section{Information Theories and Technologies for Measuring Rare Events}\label{section2}
Information measures play important roles in not only traditional information theory but also numerous applications of big data, such as detection, classification and clustering \cite{ref_Amplifying-inter-message-distance,ref_Linear-complexity-exponentially-consistent}.
In fact, by facilitating the small probability elements, some information measures focusing on rare events are proposed to settle the big data problems such as anomaly detection, feature selection and pattern recognition
\cite{
ref_Shilling-attack-detection,
ref_A-recommendation-engine,
ref_Hybrid-collaborative}.
{In these cases, rare events can be extremely eye-catching, in good agreement with the fact that the vital part of the information attracts more attention than the perfect information.
Consequently, in this paper, we merely focus on the cases where small probability events, referred to as rare events, contain importance of information.}

To characterize the importance of rare events mathematically, Message Importance Measure (MIM) \cite{ref_Message-Importance-Measure-and-Its-Application-to-Minority-Subset-Detection-in-Big-Data,
ref_Recognizing-Information-Feature-Variation,Matching-users-preference,
ref_Differential-message-importance-measure,ref_Minor-probability-events}, fixed-parameter MIM \cite{ref_Focusing-on-a-Probability-Element} and NMIM (Non-parametric MIM) \cite{ref_Non-parametric-Message-Important-Measure} are proposed, whose details are summarized in
the Table \ref{table.measure}.
We also analyze the characteristics of these information measures and compare their similarities and differences as follows.
\begin{table}[htb]%[htbp]%[!t]
\renewcommand{\arraystretch}{1.2}
% if using array.sty, it might be a good idea to tweak the value of
% \extrarowheight as needed to properly center the text within the cells
\caption{{Summary of information measures for rare events}}
\label{table.measure}
\newcommand{\tabincell}[2]{\begin{tabular}{@{}#1@{}}#2\end{tabular}}
\centering
%%%%%%%%%\resizebox{\textwidth}{56.5mm}{
% Some packages, such as MDW tools, offer better commands for making tables
% than the plain LaTeX2e tabular which is used here.
\begin{tabular}{|l|l|l|l|}%{|p{1cm}|p{2cm}|p{3.5cm}|p{3.5cm}|}%{|c|c|c|}
\bottomrule
\rowcolor[gray]{.9}
\textbf{Metrics} & \textbf{Definition} & \tabincell{c}{\textbf{Main Properties}}
& \tabincell{c}{\textbf{Key Points}}\\
%%estimation of distribution
\hline
\tabincell{l}{
MIM with \\
parameter $\varpi$\\
\cite{ref_Message-Importance-Measure-and-Its-Application-to-Minority-Subset-Detection-in-Big-Data,
ref_Recognizing-Information-Feature-Variation,Matching-users-preference,
ref_Differential-message-importance-measure,ref_Minor-probability-events}
}
&
\tabincell{l} {
$L(\textbf{\textsl{p} },\varpi)$\\
$= \log \sum_{i=1}^n p_ie^{\varpi\left(1-p_i\right)},$\\
with the parameter\\
$\varpi \ge 0$. \\
$[$
Definition 1. in  \cite{ref_Message-Importance-Measure-and-Its-Application-to-Minority-Subset-Detection-in-Big-Data}
$]$
}
&
\tabincell{l} {
$\bullet$ $L(\textbf{\textsl{p} },\varpi) \ge 0$ with $\varpi \ge 0$.\\
$\bullet$ $L(\textbf{\textsl{p} },\varpi) \ge \varpi(1-\sum p_{i}^2)$.\\
$\bullet$ $L(\alpha \textbf{\textsl{p} } + (1-\alpha)\textbf{\textsl{q} },\varpi) \ge
L(\alpha \textbf{\textsl{p} } ,\varpi)$ \\
$ \qquad \qquad + L((1-\alpha)\textbf{\textsl{q} },\varpi)$,
$\alpha \in (0,1)$.\\
$\bullet$ Event decomposition and merging: \\
          \quad The MIM can be increased by dividing\\
          \quad one event into two sub-events.
}
&
\tabincell{l} {
$\bullet$ Minority detection combined with\\
          \quad the Bayes method.\\
$\bullet$ Measure system states with small\\
          \quad probability events.\\
$\bullet$ Characterize users' preference \\
          \quad distribution for recommendation.\\
}
\\
%%Estimation of Information Measures
\hline
\tabincell{l} {
The fixed-\\
parameter \\
MIM \cite{ref_Focusing-on-a-Probability-Element}
}
&
\tabincell{l} {
$L_{j}\left (\textbf{\textsl{p}} , \varpi_{j} \right )$\\
$= \log \sum_{i=1}^n p_ie^{\varpi_{j}\left(1-p_i\right)},$\\
with the parameter\\
selection given by\\
$\varpi_{j}= F\left (p_j \right ).$\\
$[$ Eq. (3) in \cite{ref_Focusing-on-a-Probability-Element} $]$
}
&
\tabincell{l} {
$\bullet$ Principal component: $p_j$ becomes the\\
          \quad principal component in the MIM with \\
          \quad $\varpi_{j}=1/ p_j$.\\
$\bullet$ $L_a(\textbf{\textsl{p} },\varpi_{ a }) < L_b(\textbf{\textsl{p} },\varpi_{ b })$
if $p_a>p_b$.\\
$\bullet$ $L_j (\textbf{\textsl{p} },\varpi_{j}) \geq
%L (\textbf{\textsl{p} },1/ p_{max} ) > L\left (\textbf{\textsl{p} },1\right )=
\log\left( \sum_{i=1}^n p_ie^{\left(-p_i\right)}\right)+1$.\\
}
&
\tabincell{l} {
$\bullet$ To focus on the probability $p_j$ by \\
          \quad using $ \varpi_{j}= F(p_j)=1/ p_j $.\\
$\bullet$ Applied to minority subset detection.\\
$\bullet$ The mean and variance can converge \\
          \quad when samples $N_i \rightarrow \infty$.\\
}
\\
%%Estimation of entropy
\hline
\tabincell{l} {
Non-\\
parametric\\
MIM\\
\cite{ref_Non-parametric-Message-Important-Measure,ref_Storage-Code-Design-and-Transmission}
}
&
\tabincell{l} {
${L}_{non}\left( {\textbf{\textsl{p}}} \right)$\\
$= \log \sum_{i = 1}^n {{p_i}{e^{{\frac{1}{{p_i}}}-1}}},$\\
where $p_i>0$. \\
$[$ Definition 1. in \cite{ref_Non-parametric-Message-Important-Measure} $]$
}
&
\tabincell{l} {
$\bullet$ ${L}_{non}\left( {\textbf{\textsl{p}}} \right) \ge 0$.\\
$\bullet$ $ {L}_{non}\left( {\textbf{\textsl{p}}} \right)\ge n - 1$\\
$\bullet$ $ {L}_{non}({\textbf{\textsl{p}}}) + {L}_{non}({\textbf{\textsl{q}}}) \le{L}_{non}({\textbf{\textsl{pq}}})$ \\
$\bullet$ {Event decomposition and merging}:\\
          \quad The NMIM can be increased by dividing\\
          \quad one event into two sub-events.\\
}
&
\tabincell{l} {
$\bullet$  Without the constraint of parameter\\
           \quad selection.\\
$\bullet$  Exponential operator rather than \\
           \quad logarithm or polynomial operator.\\
$\bullet$  Storage code design and transmission\\
           \quad planning in wireless communication.\\
}
\\
\toprule
\end{tabular}%}
\end{table}
\begin{figure*}[htb]
\centering
\includegraphics[width=6.5in]{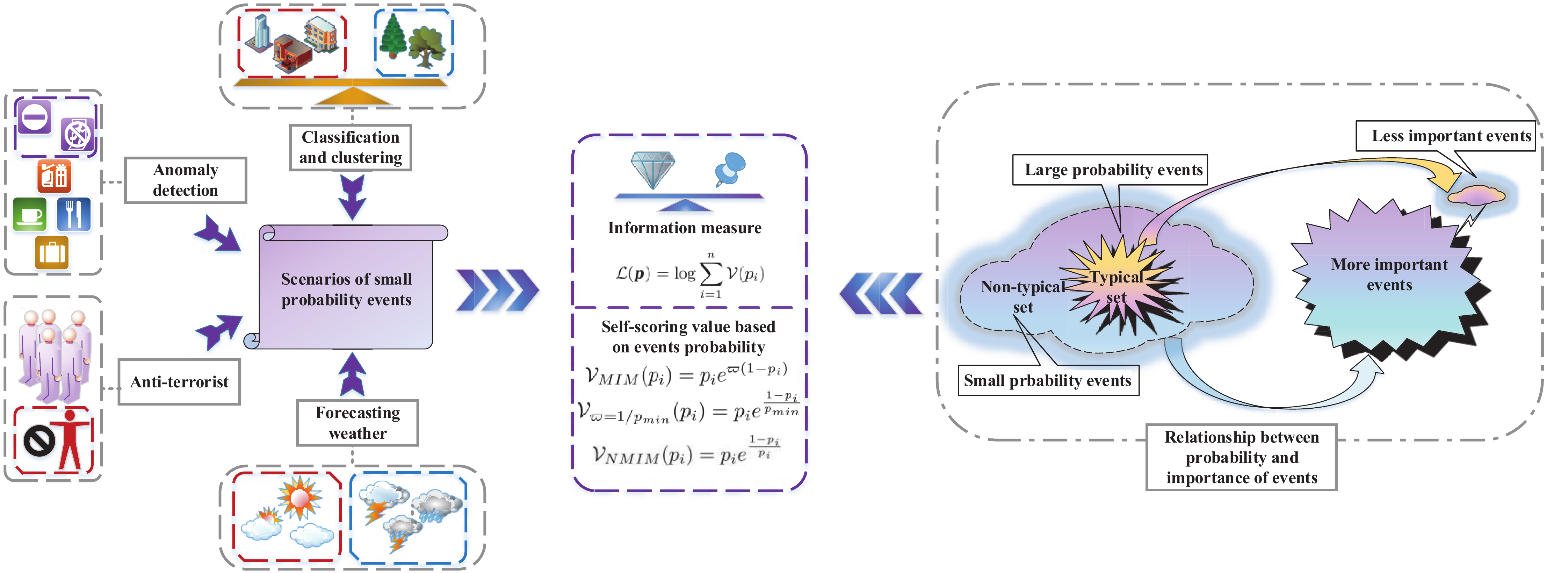}
\caption{{The interpretation of information importance measures focusing on rare events.}}
\label{fig_1}
\end{figure*}

%\vspace{-5mm}
%\begin{shaded}
%\vspace{-3mm}
%{\bf Remark 1.}
{ \em i) Intrinsic sense of the information measures:}
The common fundamental form for the information measures (including MIM, fixed-parameter MIM and NMIM) can be given by
\begin{equation}
\begin{aligned}
    \mathcal{L}( \textbf{\textsl{p}} ) = \log\sum_{i=1}^n \mathcal{V}(p_i),
\end{aligned}
\end{equation}
where $\textbf{\textsl{p}}$ is the given distribution which satisfies $\textbf{\textsl{p}}=( p_1, p_2,..., p_n)$, and the components
$\mathcal{V}(p_i)$ of MIM, fixed-parameter MIM and NMIM are respectively given by
\begin{subequations}
\begin{align}\label{aa}
    \mathcal{V}_{MIM}(p_i) = p_i e^{\varpi(1-p_i)}, \quad
\end{align}
\begin{align}
    \mathcal{V}_{\varpi=1/p_{min}}(p_i) = p_i e^{\frac{1-p_i}{p_{min}}},
\end{align}
\begin{align}
    \mathcal{V}_{NMIM}(p_i) = p_i e^{\frac{1-p_i}{p_i}}, \quad
\end{align}
\end{subequations}
where $\varpi$ denotes the coefficient of importance. Actually, these values are just the same as the intuitive notion of importance value, which can be viewed as the invariant of system, referred to as \textit{self-scoring value}.
It implies that larger weights are allocated to the small probability events than those with large probability.
Furthermore, Fig. \ref{fig_1} is shown to describe the above information measures visually.
Specifically, by treating important events from the probabilistic viewpoint, the status of the atypical sets with small probability is highlighted, which can match many scenarios such as anomaly detection, anti-terrorist activities, forecasting abnormal weather, classification and clustering for binary events.
%\end{shaded}

\begin{figure}[htb]
\centering
\includegraphics[width=4.8in]{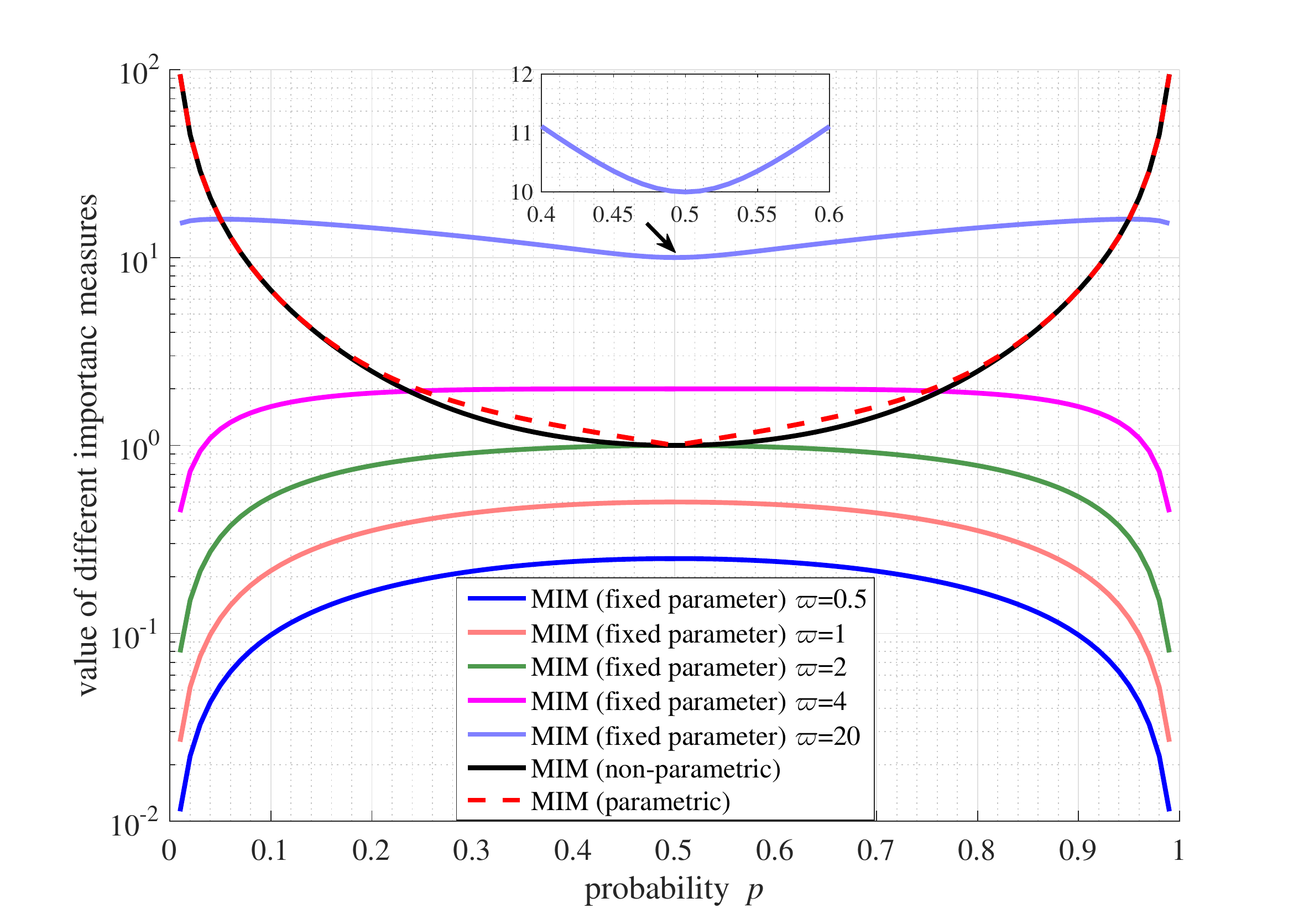}
\caption{ The comparison of different message importance measures with respect to the Bernoulli distribution ($p$, $1-p$). }
\label{fig_3MIM}
\end{figure}

%\vspace{-8mm}
%\begin{shaded}
%\vspace{-3mm}
{\em ii) Comparison of the information measures in the Bernoulli case:}
Here, the comparison of some different importance measures with respect to the Bernoulli distribution ($p$, $1-p$) is shown in Fig. \ref{fig_3MIM}.
%\end{shaded}

%\vspace{-5mm}
%\begin{shaded}
%\vspace{-3mm}
It illustrates that the parameters of MIM can make great differences on the characterization for the Bernoulli distribution. While, the non-parametric MIM (namely NMIM) and the parametric MIM (namely fixed-parameter MIM) both have similar performance on measuring small probability elements. In brief, the details of comparison are listed as follows.
\begin{itemize}
\item Due to the fact that the MIM can be influenced by parameter $\varpi$, there is no worry about beyond the computing ability of computers.
\item If the probability elements are small enough, MIM amplifies small probability not as greatly as NMIM and parametric MIM.
\item In the adjacent region of uniform distribution, the parametric MIM can perform better to amplify the smaller probability than NMIM.
\end{itemize}
%\end{shaded}

\section{Applications in Message Processing}\label{section3}

With respect to big data in IoTs, it is significant to design efficient strategies for message
processing including information compression and transmission \cite{ref_Big-data-related-technologies}.
In particular, considering rapidly exploring data \cite{ref_Spatiotemporal-stochastic-modeling}, we never need to store the whole data samples as before.
Besides, since data traffic is exponentially increasing, it is a challenge for transmission resources (including links or networks) to carry so many data pockets \cite{ref_Wireless-communications-in-the-era}.
Hence, the data processing techniques about lossy compression and transmission are investigated in many aspects \cite{ref_Lossy-compression-for-compute-and-forward,
ref_Distributed-distortion-optimization-for,
ref_Block-and-sliding-block-lossy,
ref_On-the-rate-distortion-function}.
In fact, information theory is a fundamental theory for data compression and transmission \cite{ref_Fifty-years-of}.
To be specific, it provides the optimal coding strategy and the tight bounds for the lossless and lossy compression \cite{ref_Elements-of-information-theory-2nd-edition}. Moreover, it also proposes information measures including relative entropy, Renyi divergence and f-divergence to guide information transmission and analysis \cite{ref_Renyi-divergence-and-kullback-leibler,ref_Information-theory-and-an-extension}.

%\vspace{-3mm}
%\begin{shaded}
%\vspace{-3mm}
From the perspective of rare events, a message processing architecture based on message importance measure is presented as Fig. \ref{fig_information_framework}, whose details are listed as follows:
%\end{shaded}
\begin{figure*}[htb]%[htbp]
\centering
\includegraphics[width=5.5in]{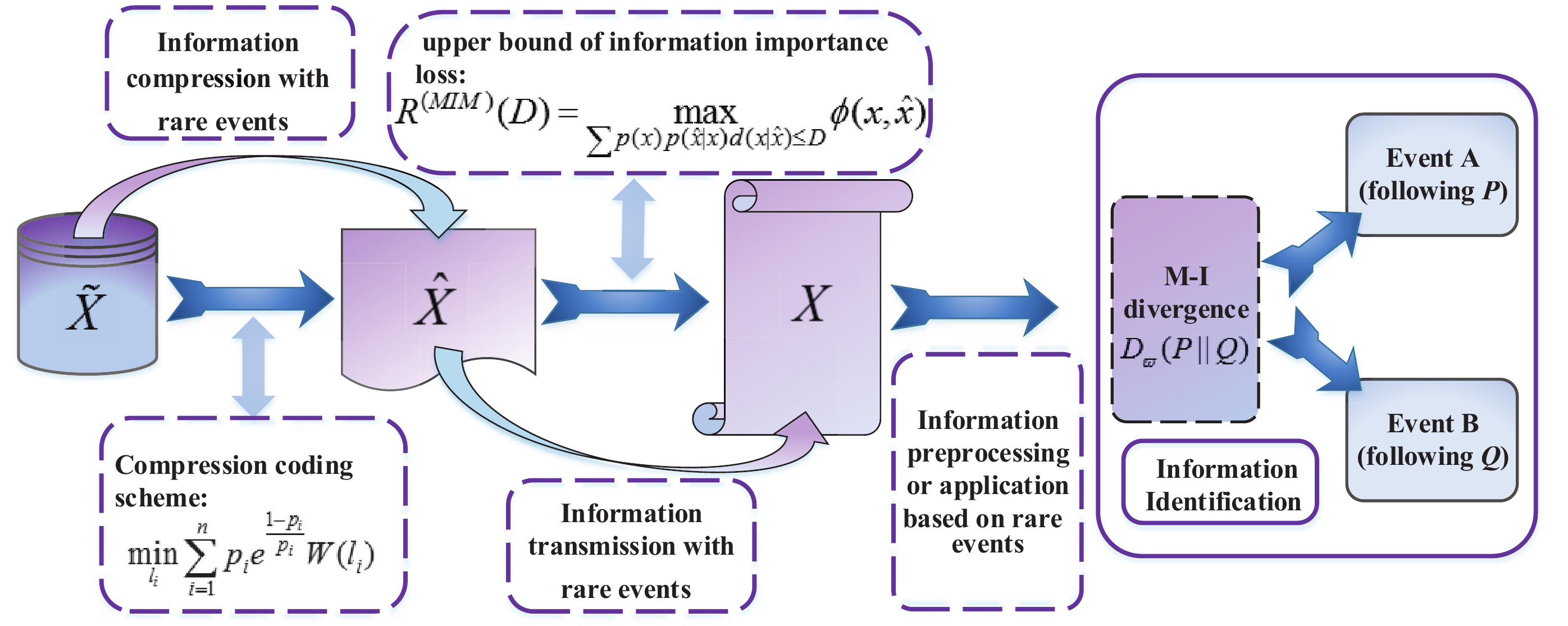}
\caption{Message processing architecture from the viewpoint of rare events.}
\label{fig_information_framework}
\end{figure*}

%\vspace{-8mm}
%\begin{shaded}
%\vspace{-3mm}
\textsl{i)}
As for the information source,
it is significant to maintain the rare events regarded as important message and lose some normal events.
In this case, it is feasible to make use of the importance measures to design lossy compression schemes.
To this end, the reconstruction error weighted by message importance can be minimized to achieve the lower bounds of code length.

\textsl{ii)}
From the viewpoint of transmission for message importance, the core idea is that the receiver can gain more amount of information from the source while maintaining the affordable loss of message importance.
In this case, the change of information measure focusing on rare events can be used to characterize the upper bound of information importance loss.

\textsl{iii)}
In the information sink, it is possible to use some information divergences to distinguish two adjacent distributions containing rare elements.
This can be regarded as a preprocessing for received data.

In terms of the specific analysis of the message processing architecture, three possible applications of information measures are summarized in the Table \ref{table.message_processing},  whose main details and interpretations are given as follows.
%\end{shaded}
\begin{table}[htb]%[htbp]%[!t]
\renewcommand{\arraystretch}{1.4}
\centering
% \extrarowheight as needed to properly center the text within the cells
\caption{{Summary of message processing applications}}
\label{table.message_processing}
\newcommand{\tabincell}[2]{\begin{tabular}{@{}#1@{}}#2\end{tabular}}
%%%%%%%%%\resizebox{\textwidth}{56.5mm}{
\begin{tabular}{|l|l|l|}%{|p{2.5cm}|p{1cm}|p{4.5cm}|}%{|c|c|c|}
\bottomrule
\rowcolor[gray]{.9}
\textbf{Work area} & \textbf{Description} & \tabincell{c}{\textbf{Key Points}}\\
%%estimation of distribution
\hline
\tabincell{l} {
Information\\ Compression \cite{ref_Non-parametric-Message-Important-Measure}
}
&
\tabincell{l} {
Compression scheme based on NMIM: \\
%\big[{l_1^*},{l_2^*},...,{l_n^*}\big]=
$\mathop {\arg\min }\limits_{{l_1},{l_2},...,{l_n}} \,\, \sum\limits_{i=1}^n {{p_i}{{e^{{1-p_i} \over {p_i}}} {W(l_i)}}}$\\
$\textrm{s.t.}\,\,\,C_{n} = \sum\limits_{i=1}^n {{l_i}},$\\
$\text{if}~l_i\le T, \quad (i=1,2,...,n),$\\
}
&
\tabincell{l} {
$\bullet$ NMIM is regarded as a measure to weight the importance\\
          \quad of code length.\\
$\bullet$ Longer codewords are allocated to the rare events rather\\
          \quad than the large probability ones.\\
$\bullet$ Lower bounds of the code length in the sense of message\\
          \quad importance.\\
$\bullet$ $W(l_i)$ is selected by different expressions such as $l_i^{-1}$ \\
          \quad and $\gamma^{-l_i}$ ($\gamma$ is a constant).\\
}
\\
%%Estimation of Information Measures
\hline
\tabincell{l} {
Information\\ Transmission \cite{ref_Storage-Code-Design-and-Transmission}
}
&
\tabincell{l} {
NMIM loss distortion function with\\
respect to distortion $D$:\\
$ {R^{(MIM)}}(D) = \mathop {\max }\limits_{ \substack{p\left( {\hat x|x} \right) \in \Omega}} \phi(X,\hat X),
$\\
$\Omega=$\\
$\{ p(\hat x|x):\sum\limits_{(x,\hat x)} {p(x)p(\hat x|x)d(x,\hat x) \le D} \},$\\
$\phi(X,\hat X) = { L_{non}({\textbf{\textsl{p}}_x}) -  L_{non}({\textbf{\textsl{p}}_{\hat x}})}$\\
}
&
\tabincell{l} {
$\bullet$ The upper bound of message importance loss caused by\\
          \quad transmission distortion.\\
$\bullet$ NMIM-loss-distortion viewpoint consisting of the message\\
          \quad importance loss $\phi(X,\hat X)$, the distortion $D$ and the \\
          \quad distribution of events.\\
}
\\
%%Estimation of entropy
\hline
\tabincell{l} {
Information\\ Preprocessing \cite{ref_Amplifying-inter-message-distance}
}
&
\tabincell{l} {
The test method for outlier detection\\ based on
information divergence $\mathcal{F}(\cdot)$:\\
$
\mathcal{F} (\hat {\textbf{\textsl{p}}}(\mathcal{X}^{(i)}); \hat {\textbf{\textsl{p}}}_{t})$\\
$\to
 \left\{
  \begin{aligned}
    \mathcal{F} (\hat {\textbf{\textsl{p}}}_{t_i}; \hat {\textbf{\textsl{p}}}_t),  & \quad \mathcal{X}^{(i)}\in \mathcal{M}_t  \\
    \mathcal{F} (\hat {\textbf{\textsl{q}}}_{f_i}; \hat {\textbf{\textsl{p}}}_t),  & \quad \mathcal{X}^{(i)}\in \mathcal{M}_f
  \end{aligned}\right.,
$
}
&
\tabincell{l} {
$\bullet$ Distribution Estimations obtained by maximum likelihood\\
          \quad estimator, k-nearest neighbor or Gaussian kernel estimator.\\
$\bullet$ Message identification divergence given by Eq. (\ref{M-I}) as $\mathcal{F}(\cdot)$ \\
          \quad to detect the outlier sequences with small probability of\\
          \quad occurrence.\\
}
\\
\toprule
\end{tabular}%}
\end{table}

\begin{figure*}[htb]
\centering
\includegraphics[width=5.5in]{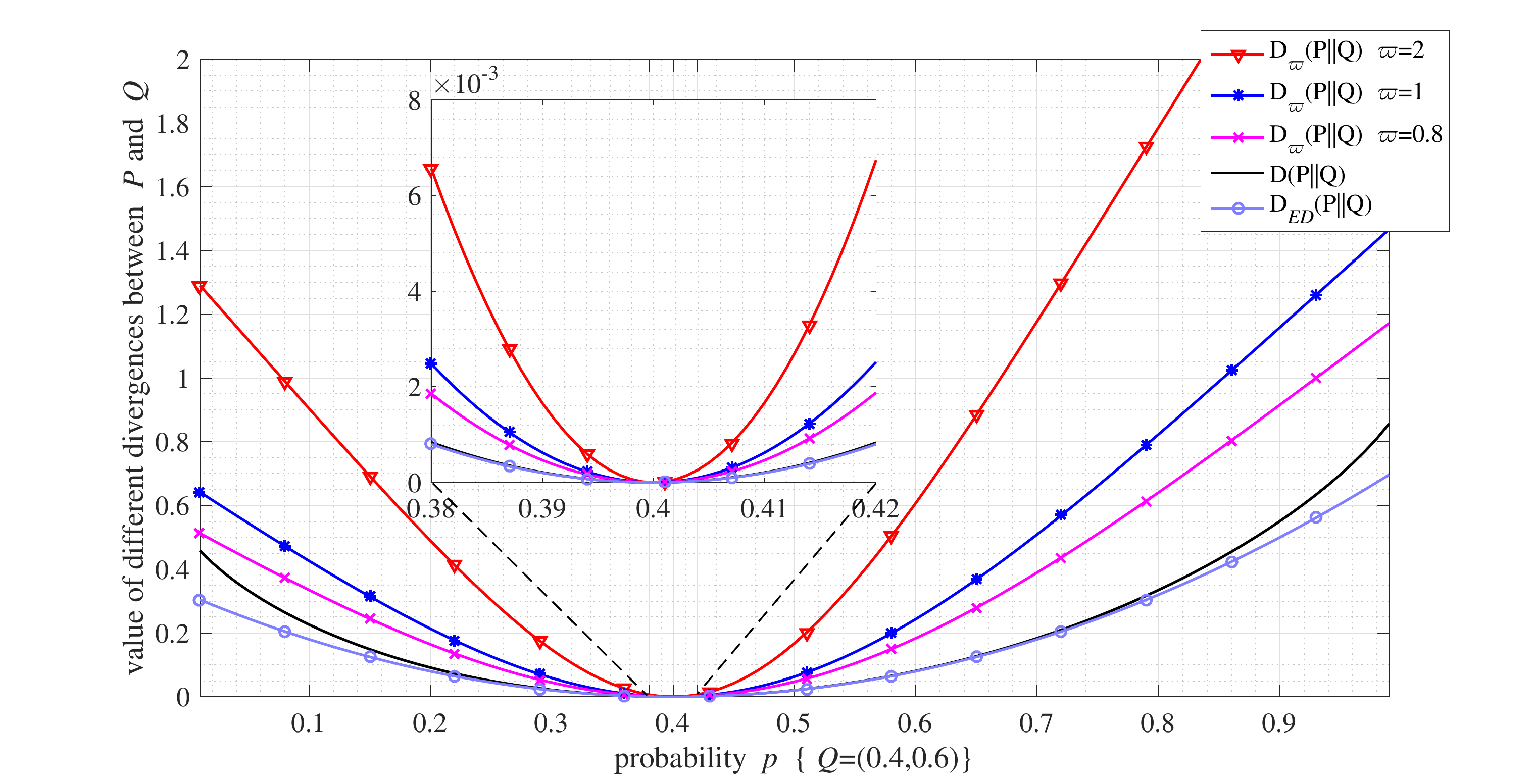}
\caption{{Comparison for different information divergences (between the probability distributions $P$ and $Q$ where $P=(p,1-p)$ and $Q=(0.4,0.6)$) including the MI divergence (with the parameter $\varpi = 2,1,0.8$), KL divergence and squared Euclidean distance.}}
\label{fig_divergence_compare}
\end{figure*}

%\vspace{-8mm}
%\begin{shaded}
%\vspace{-3mm}
\begin{itemize}
\item {\em Information Compression:}
Although standard compressions are proposed to reduce some redundant information in some degree \cite{ref_Elements-of-information-theory-2nd-edition}, there still exists large size of data that contains some unimportant message.
Further compression is considered to abandon the less vital message based on the probability of events, which may be achieved by using the compression scheme based on NMIM \cite{ref_Non-parametric-Message-Important-Measure}. In this case, lower bounds of the code length $l_i$ (with the limited total code length $C_{n}$) is obtained in the sense of message importance (based on the function of reconstruction error per unit importance, denoted by $W(l_i)$).

\item {\em Information Transmission:}
As far as big data is concerned, the dominant part of message with more importance is more favored rather than the redundant message.
%%-----------------------------------
In the traditional information transmission, some distortions or errors
may have more disastrous impacts on the important messages than worthless ones.
For instance, based on this characteristic, the strategy of unequal error protection (UEP) codes has been proposed as a reliable transmission approach \cite{ref_On-linear-unequal-error,
ref_Expanding-window-fountain,
ref_Unequal-error-protection}.
From a new viewpoint of rare events, data transmission with the constraint of message importance loss is discussed to guide the design of information transmission \cite{ref_Storage-Code-Design-and-Transmission}. In particular, the upper bound of message importance loss $\phi(X,\hat X)$ (based on NMIM operator $L_{non}(\cdot)$) is given when
there exists a kind of distortion $d(x,{\hat x})$ (such as Hamming distortion) between a source $X$ and a distortion source ${\hat X}$.

\item  {\em Information Preprocessing:}
Considering the information preprocessing, information divergences play vital roles in discriminating different distributions (namely information identification).
That is, the information divergence can be used as a test tool for outlier detection \cite{ref_Amplifying-inter-message-distance,ref_Universal-outlier-hypothesis-testin,
ref_Linear-complexity-exponentially-consistent}.
In particular, an information divergence between two distributions, denoted by $\mathcal{F}(\cdot)$, can classify the pending sample sequences
$\mathcal{X}^{(i)}$ into the normal sequence set $\mathcal{M}_t$ or the outlier sequence set $\mathcal{M}_f$.
In fact, the message identification (MI) divergence has its advantage on outlier detection \cite{ref_Amplifying-inter-message-distance},
whose definition is given by
\begin{equation}\label{M-I}
   %D_{MI}(P\parallel Q,\varpi)=
   D_{\varpi}({\textbf{\textsl{p}}}\parallel {\textbf{\textsl{q}}})=\log \sum_{i=1}^n{p_{i}e^{\left( \varpi \frac{p_i}{q_i} \right) } }-\varpi,
\end{equation}
where the adjustable coefficient $\varpi$ is positive, as well as ${\textbf{\textsl{p}}}$ and ${\textbf{\textsl{q}}}$ are two finite probability distributions in the same support set.
Here, we also take two Bernoulli distributions $P$ and $Q$ as examples to compare different information divergences shown in Fig. \ref{fig_divergence_compare}. It is illustrated that MI divergence described in the Eq. (\ref{M-I}) is more sensitive to distinguishing two distributions than the Kullback-Leibler (KL) divergence and the squared Euclidean distance when the distribution $P$ is closed to the distribution $Q$ \cite{ref_Amplifying-inter-message-distance}.
\end{itemize}
%\end{shaded}

%\vspace{-8mm}
%\begin{shaded}
%\vspace{-3mm}
{\bf Remark 1.}
{ \em
i) For information compression:
As for the data compression based on information measures for rare events, the common core idea is that the code length mainly depends on the message importance of events.
That is, the code size is mostly assigned to the small probability events.
In this case, it is applicable to use a smaller part of storage to save much more important information.}
{\em
ii) For information transmission:
%In the process of information transmission, it is worth considering to measure message importance for source compression or channel encoding.
Compared with traditional communication, the transmission for big data has its own characteristics such as larger volume of data, a wide variety of events, and the value of information.
Thus, it is sensible to preserve more information importance while reducing redundant information.
In fact, the NMIM can be used as an efficient information importance measure to design rules for communication systems.}
{\em
iii) For information preprocessing:
As for the information preprocessing,
it is possible to analyze the performance of different divergences on distinguishing distinct distributions.
Particularly, the MI divergence is a superior divergence in discerning a typical distribution from its adjacent distributions caused by rare events.}
%\end{shaded}

\section{ Applications in Data Analytics of IoTs}\label{section4}

In the view of rare events analytics of IoTs, it is required to reduce the dimension as well as estimate the distributions and their functionals efficiently. That is, we should take methods to save more computing resources and improve the efficiency of data utilization \cite{ref_Cost-Sensitive-encoding,
ref_Discriminant-analysis-based-dimension,
ref_Local-deep-feature}.
%Some techniques are investigated to efficiently exploit multidimensional samples.
Moreover, it is also necessary to analyze the relationships among rare events so that we can dig out more valuable information \cite{ref_A-Survey-of-Data-Mining,
ref_A-Deep-Learning-Approach,ref_Toward-a-Gaussian}.
From the perspective of information theory, some approaches are discussed to deal with numerous information sources and do some data mining.
Considering the relationship between information theory and big data analytics, we design an architecture based on information measures for rare events as shown in Fig. \ref{fig_Architecture_processing} whose details are summarized as follows:
\begin{figure*}[htb]
\centering
\includegraphics[width=5.7in]{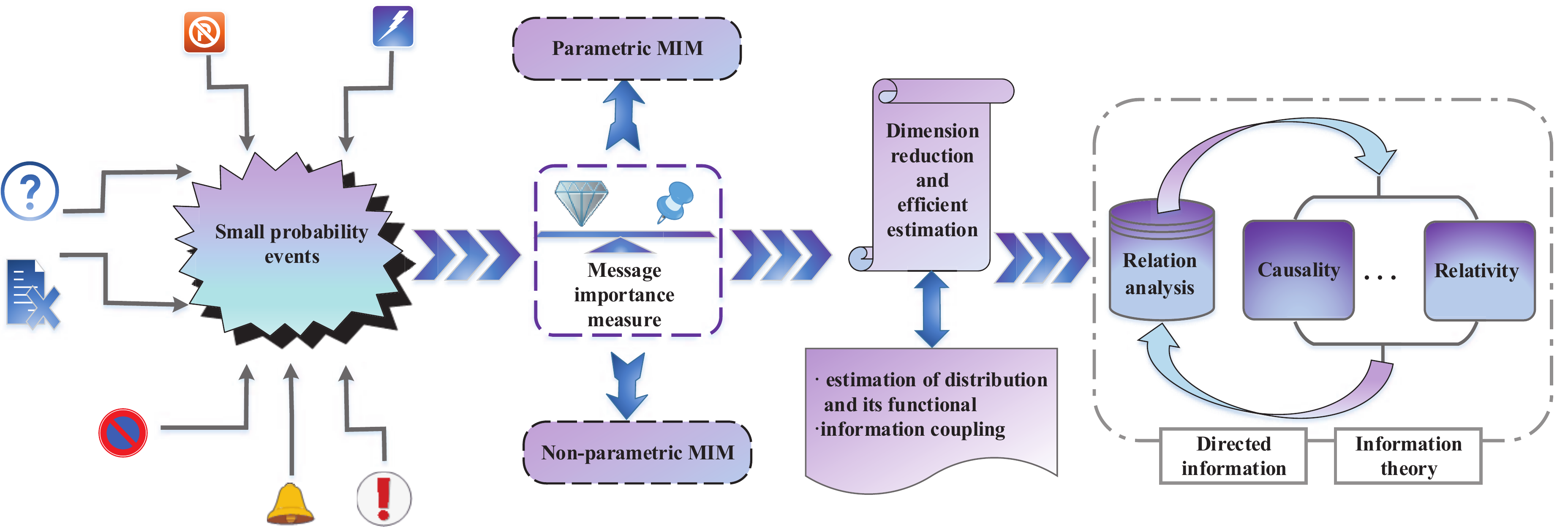}
\caption{Architecture of data analytics based on message importance of rare events.}
\label{fig_Architecture_processing}
\end{figure*}

\textsl{i) Focusing on rare events:} Rare events with small probability may contain more valuable information in some applications such as outlier detection and emergency alarm.
In this case, it is necessary to define the rare events in a specific scenario at the first step.

\textsl{ii) Selecting an information measure:} An appropriate information measure can be adopted to characterize the distribution and highlight the importance of rare events.
This is a mathematical representation of small probability events in the sense of the message importance.

\textsl{iii) Dimension reduction and efficient estimation:}
As for the sample processing, it is essential to extract the most significant information with low dimension from the original data with high dimension. Especially, in the case of rare events, we can use low dimension samples and estimate the selected information measure to decrease the computation complexity.

\textsl{iv) Analyzing relationships:}
As for big data processing, it may be efficient to analyze the relationships among rare events by use of information measures.

In the architecture of data analytics for rare events, the information measures are discussed in the Section \ref{section2}. We shall specifically introduce some applications about how to use information measures in big data analytics as follows.

\subsection{Efficient Estimation of Information Measures}\label{Efficient_Estimation}
From the perspective of big data, it is quite essential to have efficient methods to estimate information measures, especially in the case of considerably large alphabet sizes.
Whereas, the conventional estimation approaches can not work well \cite{ref_A-characterization-of-limiting,ref_Local-asymptotic-minimax,
ref_Asymptotic-Methods-in-Statistical,ref_Estimating-entropy-on-m-bins},
since that the rare events can not be observed accurately when the sample number is not very large.
It is also worth investigating asymptotics with high dimension, especially when the number of samples is not much larger than the dimension. As a result, here lists some related works in the Table \ref{table.estimation} whose details are described as follows.
\begin{table}[htb]%[!t]
\renewcommand{\arraystretch}{1.2}
%%%%%\resizebox{\textwidth}{40mm}{
% if using array.sty, it might be a good idea to tweak the value of
% \extrarowheight as needed to properly center the text within the cells
\caption{Summary of literature on the efficient estimations of distribution and its functional}
\label{table.estimation}
\newcommand{\tabincell}[2]{\begin{tabular}{@{}#1@{}}#2\end{tabular}}
\centering
% Some packages, such as MDW tools, offer better commands for making tables
% than the plain LaTeX2e tabular which is used here.
\begin{tabular}{|l|l|l|}%{|p{2.5cm}|p{1cm}|p{4.5cm}|}%{|c|c|c|}
\bottomrule
\rowcolor{mygray}
\textbf{Work Area} & \textbf{Related Work} & \tabincell{c}{\textbf{Key Points}}\\
%%estimation of distribution
\hline
Estimation of Distribution & \cite{ref_Minimax-estimation-of-discrete},
\cite{ref_Minimax-estimation-of-discrete-distributions-under},
\cite{ref_Minimax-risk-over},
\cite{ref_Ideal-spatial-adaptation},
\cite{ref_Beyond-histograms} &
\tabincell{l} {
$\bullet$ Minimax estimation under $\ell_1$ loss \\
$\bullet$ Maximum likelihood estimator (MLE) based on\\ risk functions \\
$\bullet$ Non-asymptotic and asymptotic upper and\\ lower bounds \\
$\bullet$ Alphabet sizes and the number of samples\\ both increasing
}
\\
%%Estimation of Information Measures
\hline
{\tabincell{l}{Estimation of Functionals of \\ Distribution}
} &
\tabincell{l} {
\cite{ref_Minimax-estimation-of-information-measures},
\cite{ref_Minimax-estimation-of-functionals-of-discrete-distributions},
\cite{ref_Mutualinformation-based-registration-of-medical-images},
\cite{ref_Alignment-by-maximization-of-mutual-information},
\cite{ref_Mutual-information-analysis},\\
%\cite{ref_Alignment-by-maximization-of-mutual-information},
%\cite{ref_Mutual-information-analysis},
\cite{ref_Maximum-Likelihood-Estimation-of-information-measures}
}
&
\tabincell{l} {
$\bullet$ Minimax estimator with the best polynomial\\ approximation\\
$\bullet$ Support size comparable with the number of\\ samples \\
$\bullet$ Non-smooth and smooth regimes for functionals
%$\bullet$ Effective sample size enlargement
}
\\
%%Estimation of entropy
\hline
Entropy Estimation & \cite{ref_Adaptive-estimation-of-Shannon-entropy},
\cite{ref_Optimal-entropy-estimation-on-large-alphabets-via-best-polynomial-approximation},
\cite{ref_Does-dirichlet-prior-smoothing},
\cite{ref_Ensemble-estimators-for-multivariate-entropy-estimation} &
\tabincell{l} {
$\bullet$ Adaptive estimation framework\\
$\bullet$ Dirichlet prior smoothing\\
$\bullet$ Ensemble of plug-in estimators with weights
}
\\
%%Estimation of Information divergence
\hline
Information Divergence Estimation &
\tabincell{l} {
\cite{ref_Estimation-of-KL-divergence-between-large-alphabet-distributions},
\cite{ref_Estimating-the-unseen},
\cite{ref_Minimax-rates-of-entropy-estimation},
\cite{ref_Divergence-estimation-of-continuous-distributions},
\cite{ref_Divergence-estimation-for-multidimensional-densities},\\
\cite{ref_Estimating-divergence-functionals-and-the-likelihood-ratio},
\cite{ref_Minimax-rate-optimal-estimation-of-KL-divergence},
\cite{ref_Ensemble-estimation-of-multivariate-f-divergence}
}
&
\tabincell{l} {
$\bullet$ Augmented plug-in estimator\\
$\bullet$ Methodology with the polynomial approximation\\ and the plug-in rule \\
$\bullet$ Optimally weighted ensemble estimation
}
\\
\toprule
\end{tabular}%}
\end{table}

\begin{itemize}
%%%%%%%%%%1
\item \textit{Estimation of Distributions}: Based on some risk functions, different distribution estimations are investigated which play crucial roles in the information measure estimation \cite{ref_Minimax-estimation-of-discrete,
ref_Minimax-estimation-of-discrete-distributions-under,ref_Minimax-risk-over,
ref_Ideal-spatial-adaptation,ref_Beyond-histograms}.
    For example, in the case that the alphabet size $S$ increases with the number of samples $n$, a minimax estimation is discussed under the $\ell_1$ loss (which is defined by
$\mathbb{E}_{P} \sum_{i=1}^{S}| p_i-\hat p_i|$).
    This estimator has better performance on non-asymptotic upper and lower bounds of risk than maximum likelihood estimator (MLE).
%%%%%%%%%%2
\item \textit{Estimation of functionals of distribution}: When the unknown support size $S$ is not smaller or even larger than the samples number $n$, a general methodology based on the minimax estimator is presented to estimate the functionals of distribution \cite{ref_Minimax-estimation-of-information-measures,
ref_Minimax-estimation-of-functionals-of-discrete-distributions}. Compared with the minimax estimator with non-smooth and smooth regions, the MLE is exactly sub-optimal in the large support \cite{ref_Mutualinformation-based-registration-of-medical-images,
ref_Alignment-by-maximization-of-mutual-information,ref_Mutual-information-analysis,
ref_Maximum-Likelihood-Estimation-of-information-measures}.
%%%%%%%%%%3
\item \textit{Entropy Estimation}: As a widely used information measure, entropy is worth estimating especially. An adaptive estimation framework is adopted to achieve the minimax rates in spite of the unknown support size $S$ of distribution \cite{ref_Adaptive-estimation-of-Shannon-entropy}. Besides, the estimator based on the best polynomial approximation also has the same performance \cite{ref_Optimal-entropy-estimation-on-large-alphabets-via-best-polynomial-approximation}. Moreover, an inferior estimator is constructed by use of Dirichlet prior smoothing, which is similar to MLE but not as good as the above two \cite{ref_Does-dirichlet-prior-smoothing}. In addition, an ensemble of plug-in estimators with weights is proposed to protect the results of estimation from decaying with the increase of sample dimension  \cite{ref_Ensemble-estimators-for-multivariate-entropy-estimation}.
%%%%%%%%%%4
\item \textit{Information Divergences Estimation}: As a class of information measures, information divergences such as KL divergence, Hellinger distance and $\ell_2$-divergence can be estimated in some similar ways \cite{ref_Estimation-of-KL-divergence-between-large-alphabet-distributions,
ref_Estimating-the-unseen,ref_Minimax-rates-of-entropy-estimation,
ref_Divergence-estimation-of-continuous-distributions,
ref_Divergence-estimation-for-multidimensional-densities,
ref_Estimating-divergence-functionals-and-the-likelihood-ratio}. In this regard, an augmented plug-in estimator and a methodology with the combination of polynomial approximation and plug-in rule are constructed to achieve the consistent estimator and the minimax rate-optimal estimator respectively \cite{ref_Minimax-rate-optimal-estimation-of-KL-divergence}.
    Moreover, an optimally weighted ensemble estimator is also designed, which has good performance in the cases of high dimension \cite{ref_Ensemble-estimation-of-multivariate-f-divergence}.
\end{itemize}

In fact, the above classifications are based on the work areas of estimation. While, there exist some common criterions which can unify these estimators \cite{ref_Minimax-estimation-of-functionals-of-discrete-distributions,
ref_Maximum-Likelihood-Estimation-of-information-measures}, whose details are discussed as follows.

\textit{i) The maximum risk:} Essentially, the MLE of distributions or their functionals complies with the maximum risk criterion which is given by
\begin{equation}\label{eq.max_risk}
\begin{aligned}
& \sup_{P\in M_{S}} \mathbb{E} \{ D_{error}(F(P) - \hat F)\} ,
\end{aligned}
\end{equation}
where $D_{error}$ denotes a kind of error metric such as the one-norm and two-norm, $F(P)$ is a function of the distribution $P$ whose support is $M_{S}$ and $\hat F$ is the estimation for $F(P)$. In general, the MLE of distributions can be regarded as the fundamental plug-in estimator which is given by
\begin{equation}
\begin{aligned}
& \hat p_i = \frac{X_i}{n}, \quad X_i= \sum_{j=1}^{n} I_{\{ Z_j = i\}},\quad (1\le i\le S),
\end{aligned}
\end{equation}
where $Z_j$ ($j \in \{1, 2, ..., n\}$) denotes the sample value, $n$ is the sample number and $S$ is the support size. Furthermore, we can substitute $\hat p_i$ into the functionals including $F(P) = P$ (namely the distribution itself) to obtain the estimation for the functionals of distribution.
Moreover, as another example of MLE, the Dirichlet prior smoothing estimator is similar to plug-in estimator in the case of maximum squared risk, which is given by
\begin{equation}
\hat P_{D} = \frac{n}{n+ \sum_{i=1}^{S}\alpha_i} \hat P +
\frac{\sum_{i=1}^{S}\alpha_i}{n+ \sum_{i=1}^{S}\alpha_i} \frac{\bm{\alpha}}{n+ \sum_{i=1}^{S}\alpha_i},
\end{equation}
where $S$ is the alphabet size, $\hat P$ is an empirical distribution, and $\bm{\alpha} = (\alpha_1, \alpha_2..., \alpha_S)$ denotes the parameter vector which is adjustable.
Besides, the ensemble of plug-in estimators with weights also belongs to MLE, which is defined by
\begin{equation}
\begin{aligned}
 \hat F_{e} = \sum_{l \in \bar l} \lambda_l \hat F_l, \quad
(\sum_{l \in \bar l} \lambda_l = 1),
\end{aligned}
\end{equation}
where $\hat F_l$ is the plug-in estimator or its function, $\bar l = \{l_1, l_2,..., l_L\}$ is a set of parameters and $\lambda_l$ denotes the weight value.
In this estimator, the weights can be adjusted by using different optimal rules flexibly.

\textit{ii) The minimax risk:} In terms of the minimax estimator for distributions or information functionals, it is based on the criterion minimizing the maximum risk of MLE which is given by
\begin{equation}
\begin{aligned}
& \inf_{\hat F} \sup_{P\in M_{S}} \mathbb{E} \{ D_{error}(F(P) - \hat F)\} ,
\end{aligned}
\end{equation}
in which the notations are the same as those in the Eq. (\ref{eq.max_risk}).
As an instance of the minimax estimator, an approach based on the polynomial approximation rule is proposed, which treats the estimation problem as two cases of ``small $p_i$'' and ``large $p_i$''  ($p_i$ denotes the probability element).
In the case of ``small $p_i$'', the best polynomial approximation is used to guide the estimation, which is given by
\begin{equation}
P^{*}_K(x) = \arg\min_{{P\in {\Psi}_K} } \max_{x\in \Omega} |g(x)-P(x)|
\end{equation}
where $g(x)$ is the objective function, ${\Psi}_K$ is the set of polynomials with order no more than $K$ on the domain $\Omega$.
Moreover, in the case of ``large $p_i$'', the estimation can be obtained by use of a kind of MLE such as the plug-in estimator.

Moreover, in order to see the reliability of the estimators based on these criterions (including the minimax risk or the maximum risk of MLE), it is necessary to compare the corresponding performance in some specific cases.
Here, the results of estimating some classical information measures are summarized in the Table \ref{tab:comparison_risk} in which $H(P)=-\sum_{i=1}^{S} p_i \ln p_i$ denotes the Shannon entropy, $H_{\xi}(P) = \sum_{i=1}^{S} p_i^{\xi} $ ($\xi > 0$) is the dominant part of Renyi entropy, $S$ is the support size, $n$ is the samples' number, and
the notation $a_{k} \lesssim b_{k}$ denotes $\sup_{k} \frac{a_{k}}{b_{k}} \le A$ ($A$ is a constant). It is remarkable that the performance of the minimax estimator with $n$ samples is equal to the MLE with $n \ln n$ samples in the case of small probability estimation, which is called ``effective sample size enlargement''.
\begin{table}[htb]
\renewcommand{\arraystretch}{1.3}
\centering
    \caption{{Performance of minimax estimator and MLE and the comparison} \cite{ref_Minimax-estimation-of-functionals-of-discrete-distributions,
ref_Maximum-Likelihood-Estimation-of-information-measures} }\label{tab:comparison_risk}
    \newcommand{\tabincell}[2]{\begin{tabular}{@{}#1@{}}#2\end{tabular}}
\begin{tabular}{|c|c|c|}
\bottomrule
\rowcolor{mygray}
\textbf{Functional of distribution} & \textbf{Minimax squared error rates} & \textbf{Maximum squared error rates of MLE}\\
\hline
$H(P)$ &
$\frac{S^2}{(n\ln n)^2}+\frac{\ln^2 n}{n}$, ($\frac{S}{\ln S}\lesssim n$)\cite{ref_Minimax-estimation-of-discrete,
ref_Minimax-estimation-of-functionals-of-discrete-distributions,ref_Estimating-the-unseen,
ref_Minimax-rates-of-entropy-estimation} &
$\frac{S^2}{n^2}+\frac{\ln^2 n}{n}$ \cite{ref_Maximum-Likelihood-Estimation-of-information-measures}\\
\hline
$H_{\xi}(P), 0< \xi \le \frac{1}{2}$ &
$\frac{S^2}{(n\ln n)^{2\xi}}$, ($\frac{S^{\frac{1}{\xi}}}{\ln S}\lesssim n$, $\ln n \lesssim \ln S$) \cite{ref_Minimax-estimation-of-discrete,
ref_Minimax-estimation-of-functionals-of-discrete-distributions}&
$\frac{S^2}{n^{2\xi}}$, ($S^{\frac{1}{\xi}} \lesssim n $) \cite{ref_Maximum-Likelihood-Estimation-of-information-measures}\\
\hline
$H_{\xi}(P), \frac{1}{2} < \xi < 1$ &
$\frac{S^2}{(n\ln n)^{2\xi}} + \frac{S^{2-2\xi}}{n}$, ($\frac{S^{\frac{1}{\xi}}}{\ln S}\lesssim n$) \cite{ref_Minimax-estimation-of-discrete,
ref_Minimax-estimation-of-functionals-of-discrete-distributions} &
$\frac{S^2}{n^{2\xi}}+ \frac{S^{2-2\xi}}{n}$, ($S^{\frac{1}{\xi}} \lesssim n $) \cite{ref_Maximum-Likelihood-Estimation-of-information-measures}\\
\hline
$H_{\xi}(P), 1< \xi < \frac{3}{2}$ &
$\frac{1}{(n\ln n)^{2-2\xi}}$, ($n \ln n \lesssim S$) \cite{ref_Minimax-estimation-of-discrete,
ref_Minimax-estimation-of-functionals-of-discrete-distributions}&
$\frac{1}{n^{2-2\xi}}$, ($ n \lesssim S$)\cite{ref_Maximum-Likelihood-Estimation-of-information-measures}\\
\hline
$H_{\xi}(P),  \frac{3}{2} \le  \xi $ & $\frac{1}{n}$ \cite{ref_Minimax-estimation-of-discrete,
ref_Minimax-estimation-of-functionals-of-discrete-distributions} &  $\frac{1}{n}$ \cite{ref_Minimax-estimation-of-discrete,
ref_Minimax-estimation-of-functionals-of-discrete-distributions,
ref_Maximum-Likelihood-Estimation-of-information-measures} \\
\toprule
\end{tabular}
\end{table}

\subsection{Dimension Reduction Based on Information Coupling}
In the era of big data, there exists a big buzz word, ``dimension reduction'', which is involved in many fields such as machine learning, data mining, computer vision, etc. In order to solve this problem, more and more new techniques are being developed including
principal component analysis, independent component analysis
and regression analysis \cite{ref_Principal-component-analysis,
ref_Independent-component-analysis,
ref_Statistical-Models-Theory-and-Practice,
ref_K-means-clustering-via-principal-component-analysis}.
Besides, lots of applicable algorithms enable these new developed approaches to be used in many applications \cite{ref_A-new-approach-to-linear-filtering,
ref_Probabilistic-Graphical-Models}.
However, these approaches are all designed from the viewpoint of the space of data rather than the intrinsic information flow.

On the contrary, the information coupling based on information measures
is discussed to construct a framework for information-centric data processing.
In fact, it is a novel view to analyze the information exchange process of relative data nodes by use of information coupling.

Mathematically, information coupling can be formulated in a fundamental communication scenario, where the input $X$ contributes to the output $Y$ through a transition probability matrix $W_{Y|X}$. In a typical communication system, a message $U$ can form a Markov chain $U \to X \to Y$ with the input $X$ and the output $Y$, where the message $U$ is encoded into the input $X$.
In order to design an efficient encoding scheme, it is usual to maximize the mutual information $I(U;Y)$ depending on the distribution $P_{U}$ and the conditional distributions $P_{X|U=u}$.
Similarly, the information coupling is to maximize the objective function $I(U;Y)$ constrained by a small mutual information $I(U;X)$. The constraints satisfy that the conditional distributions $P_{X|U}(\cdot|u)$ are neighbors of the marginal distribution $P_X$.
That is, the information coupling \cite{ref_Linear-information-coupling-problems} can be given by
\begin{flalign}\label{equ:couple_optization}
\mathop {\max }\limits_{U \to X \to Y}\,\,\, & \frac{1}{n} I(U;Y),  \\
\textrm{s.t.}\,\,\,& \frac{1}{n} I(U;X) \le \sigma, \tag{\theequation a}\label{equ:couple_condition a}\\
& \frac{1}{n} || P_{X|U=u} - P_X ||^2= O(\sigma), \forall u, \tag{\theequation b}\label{equ:couple_condition b}
\end{flalign}
where the parameter $\sigma$ is small enough.

In practice, the solution of the optimization problem about information coupling can provide a theoretical optimal result for dimension reduction from the perspective of information correlation.
{This can guide us to approximate the optimum by using low-dimensional information to represent the high-dimensional data.
Specifically, suppose that there exists a hidden source sequence ${\bm x}_n =\{x_1, x_2,...,x_n\}$ following the distribution $P_X$, an observed sequence ${\bm y}_n = \{y_1, y_2,...,y_n\}$ following the distribution $P_Y$, and a transfer matrix $W_{Y|X}$ between the input $X$ and output $Y$.
In order to infer the hidden source $X$ from $Y$, we usually require a sufficient statistic of ${\bm y}_n$ containing the whole information of ${\bm x}_n$.
While, it is difficult to compute the statistic in the cases of the high dimensional structures of ${\bm x}_n$ and ${\bm y}_n$.
To reduce the dimension, we would like to acquire a statistic from the observation ${\bm y}_n$ to characterize a certain feature of ${\bm x}_n$. According to the information coupling, a feature $U$ in ${\bm x}_n$ is the most efficiently extracted from the observed data ${\bm y}_n$ in terms of the maximized mutual information $I(U;Y)$, which corresponds to the solution of this optimization problem.
This efficient statistic based on the feature $U$, can be considered as a low-dimensional label containing the most significant information of the high-dimensional data, which implies an information theoretic method to reduce dimension} \cite{ref_The-Linear-information}.
%Shao-Lun Huang, Lizhong Zheng
%The Linear Information Coupling Problems
%https://arxiv.org/abs/1406.2834, pp. 1--27
%Submitted on 11 Jun 2014

{\bf Remark 2.}
{\em
Actually, it is not difficult to see that the information coupling is an efficient tool for statistics, which can extract the significant information from high dimensional original data.
This can correspond to the goal of the dimension reduction and feature extraction for the rare events, which may use $\phi(U;X)   ={ \mathcal{L}({\textbf{\textsl{p}}_U}) -  \mathcal{L} ({\textbf{\textsl{p}}_{X}})}$ to replace $I(U;X)$ to take the message importance transfer quantifying.}

\subsection{ Directed Information for Relationship Analysis }
Directed information derived from information theory seems to be a commonly used approach, which can identify the interplay and causality between two stochastic processes \cite{ref_Universal-estimation-of-directed-information,
ref_Investigating-causal-relations,
ref_Using-directed-information-to-build-biologically,
ref_Universal-divergence-estimation,ref_Thecontext-tree-weight-ingmethod,ref_Universal-directed-information}.
Furthermore, it is also rational to adopt this approach to analyze the stochastic processes with rare events.
Some details of directed information are given as follows.

In order to solve the causality problem in information systems \cite{ref_Causality-feedback-and-directed,ref_The-bidirectional-communication-theory}, an information measure, referred to as ``directed information'', is defined as
\begin{equation}
    I(X^n \to Y^n) = \sum_{i=1}^n I(X^i; Y_i|Y^{i-1}),
\end{equation}
where $X^{n}=(X_1,X_2,\ldots, X_n)$ and $Y^{n}=(Y_1,Y_2,\ldots, Y_n)$ are independently random sequences, while $X_i$ and $Y_i$ ($i=1 ,2 ,..., n$) are random variables, and $I(\cdot)$ denotes the mutual information.
Moreover, due to the fact that the upper bound of the feedback channel capacity can be obtained by maximizing the normalized directed information \cite{ref_Interpretations-of-directed-information,ref_Directed-Information-for-Channels-With-Feedback}, another formulation of directed information is given by
\begin{equation}
    I(X^n \to Y^n) = \sum_{i=1}^n I(X_i; Y_i^n | X^{i-1}, Y^{i-1}).
\end{equation}
which is obtained by use of the slide information ($X^{i-1}, Y^{i-1}$) \cite{ref_A-coding-theorem}.

Furthermore, this information measure has been adopted in some applications of relationship analysis, such as the computational biology with intrinsic causality \cite{ref_Estimating-the-directed-information,ref_Mapping-information-flow}, the prediction of rate distortion \cite{ref_MAchieving-the-Gaussian-rate} and the data compression with causal side information.
Besides, the directed information provides an upper bound for the growth rates of optimal portfolios, which can also tightly bound the horse race gambling \cite{ref_Universal-estimation-of-directed-information}.
Notably, directed information can also measure the best error exponent for hypothesis testing which may be involved with the rare events identification.
%This can help to understand the physical meaning, namely, the amount of information that causally transfers from one random sequence to the other one.

{\bf Remark 3.}
{\em
Directed information is an efficient information measure which can interpret the causality transfer between two variables.
Actually, this measure provides a significant tool to analyze the causal side information. Besides, it also plays an crucial role in dealing with the inference problem involved with causal influence factors. Similar method for the extension of MIM is necessary, which may bring some new insights on the massage importance discussion.}

\begin{figure*}[htb]
\centering
\includegraphics[width=6.5in]{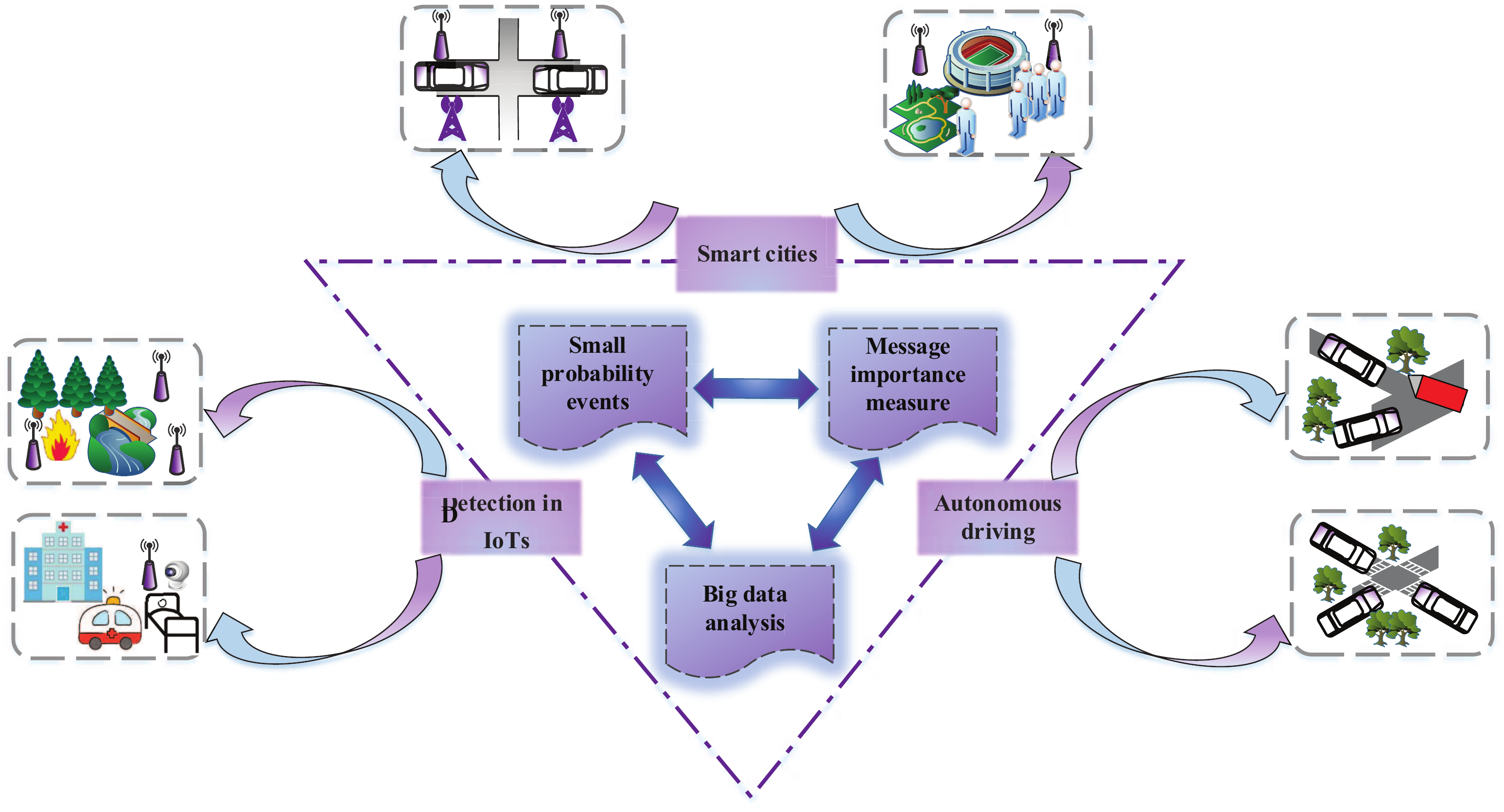}
\caption{Architecture of small probability events (namely rare events) processing based on information measures for challenging applications.}
\label{fig_Architecture_future}
\end{figure*}
%\vspace{-3mm}
%\begin{shaded}
%\vspace{-3mm}
\subsection{Rare Events Detection for Probability Derivation Process}
In the data mining of IoTs, some scenarios such as urban abnormal pattern recognition as well as fire early warning and detection, can be treated as probability derivation processes which may be characterized and analyzed by means of information theory.
It is worth noting that rare events detection lies in the intersection of the probability derivation process and the practical applications related to information measures. This problem has been investigated from many perspectives.
In particular, the common methods of rare events detection are proposed based on the specific models or frameworks \cite{ref_A-unified-framework-for-event,ref_Semi-Markov-switching,
ref_Spatio-temporal-network-anomaly,ref_Group-anomaly-detection,ref_Detecting-and-classifying}, such as Bayesian network anomaly detection, anomaly pattern classification in images, as well as normal behaviors definition for data points or groups.

As a typical probability derivation process, urban abnormal events detection is investigated widely, which may provide advices for governments and communities in smart city planning and management.
In this regard, spatio-temporal data or multiple data sources are used to detect rare events of urban traffic states, such as mining uncommon trajectory of people, detecting road traffic anomalies \cite{ref_On-mining-anomalous}, as well as identifying anomalous regions or locations
\cite{ref_Detecting-collective-anomalies,ref_Spatio-Temporal-Event-Detection-Using,
ref_Anomalous-event-detection-on-large-scale}.
The essential idea of these approaches is to construct a conditional probability model based on Hidden Markov process or Maximum Likelihood rule to detect or predict anomalous events. That is, the underlying distribution of rare patterns can be obtained in the probabilistic models which are constructed based on the different patterns of spatio-temporal data.

Moreover, message measures based on similarity and correlation also play crucial roles in identifying urban abnormal events \cite{ref_Multicriteria-Similarity-Based,
ref_Visual-Analysis-of-Time-Series}.
For instance, $L_{-\infty}$ distance is adopted as a kind of similarity measurement to evaluate the degree of anomalous traffic \cite{ref_Temporal-outlier-detection}.
Besides, KL divergence is also commonly used as a metric to measure correlation \cite{ref_Detecting-rare-events-using-Kullback¨CLeibler,
ref_Anomaly-Detection-and-Localization}.
In video surveillance systems of urban traffic states, when a small video clip is represented as a histogram of multi-set bag of codewords by using Fourier based trajectory feature descriptor \cite{ref_Detecting-rare-events-using-Kullback¨CLeibler},
KL divergence is applied to classify the pending video clips into the normal or abnormal ones. The corresponding metric based on KL divergence is given by
\begin{equation}
\begin{aligned}
    D_{KL}(Q||P_{0})-D_{KL}(Q||P_{1}) = \ln \prod_{i=1}^{K-1} \Big(\frac{p(v_i|c=1)}{p(v_i|c=0)}\Big)^{q_i},
\end{aligned}
\end{equation}
where $D_{KL}(\cdot)$ denotes the operator of KL divergence, $p(v_i|c=1)$ and $p(v_i|c=0)$ are probability elements from the codewords of normal video clips and abnormal ones (the corresponding distributions are $P_1$ and $P_0$), $q_i$ denotes the probability element from the codewords of pending video clips (the corresponding distribution is $Q$).
Furthermore, a spatio-temporal detector for the mixture of dynamic textures (MDT) model is proposed, in which the center-surround saliency detection is based on the KL divergence between feature responses and events class labels \cite{ref_Anomaly-Detection-and-Localization}:
\begin{equation}
\begin{aligned}
    &D_{KL}(P_{{\bm X}|c}||P_{\bm X})  \\
    & \qquad  \doteq \sum_{i}
    \Big\{ \pi_{i}^{c} \log
    \frac{\sum_{j}^{K_c} \pi_{j}^{c} {e}^{(-D_{KL}(p^{i}_{{\bm X}|c}||p^{j}_{{\bm X}|c}))}}
    {\sum_{j}^{K_0+K_1} \omega_{j} {e}^{(-D_{KL}(p^{i}_{{\bm X}|c}||p^{j}_{{\bm X}}))}} \Big\},
\end{aligned}
\end{equation}
where $p^{i}_{{\bm X}|c}$ are class-conditional densities (based on the class $c\in \{0,1\}$), $p^{j}_{{\bm X}}$ are sample densities, $\pi_{j}^{c}$ and $\omega_{j}$ are parameters, $K_c$ ($c\in\{0,1\}$) denotes the number of samples in the corresponding class $c$.

Similar to the KL divergence, the message measures mentioned in Section \ref{section2} may be
also efficient in rare events detection for spatio-temporal data and may perform better in some special data sets, which can be investigated further in probability derivation processes.

{\bf Remark 4.}
{\em
Some message measures reveal the similarity or correlation for probability derivation processes.  Specifically, these measures can be regarded as criteria for urban abnormal events mining.
In general, it is promising to make good use of novel information measures to extend the strategies of rare events detection.}
%\end{shaded}

\section{Future Challenges}\label{section5}
{Considering future research directions, new approaches and challenging applications can promote the development of information measures with respect to rare events.}
By combining big data analytics,
an architecture of rare events processing based on information measures is constructed shown in Fig. \ref{fig_Architecture_future}.
{In particular, we can apply big data analytics and information measures in the challenging scenarios involved with rare events, including smart cities, autonomous driving, and detection in
IoTs.}

\begin{figure*}[htb]
\centering
\includegraphics[width=6.5in]{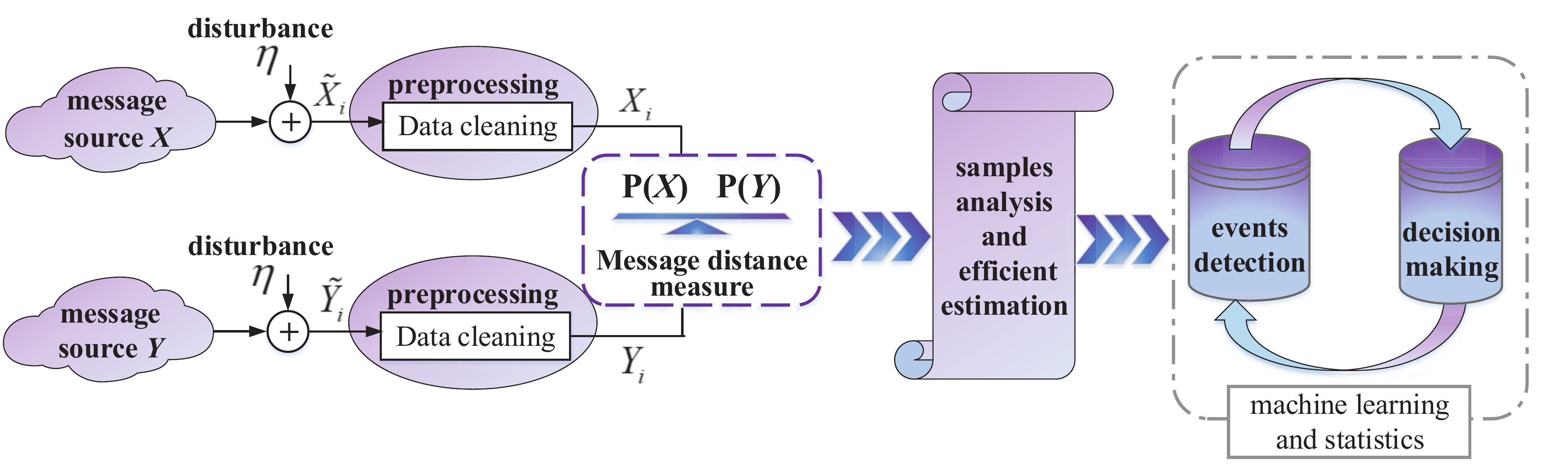}
\caption{The framework for rare events detection. }
\label{fig_detection_future}
\end{figure*}

Actually, in the above applications, the common technique playing a core role is rare events detection. Here, we design a technology framework in the viewpoint of information measures to help to detect rare events as shown in Fig. \ref{fig_detection_future}.
To be specific, assume there exist two different kinds of message sequences in the data set, that is, the data set consists of two message sources $X$ and $Y$ with different distributions. In this case, the message sequences from the message source $Y$ are considered as the rare events. The goal of our framework is to detect message sequences of $Y$. Our core idea is to make use of information measures such as KL divergence, Renyi divergence and f-divergence to identify the two kinds of information distributions.
In this case, we assume that how to design efficient information measures is a fundamental problem in the first step.
Moreover, when an information measure is obtained, we also need to analyze the samples in the message sequences and take efficient methods to estimate the information measure. Furthermore,
it is applicable to classify estimated results by resorting to the machine learning algorithms so that we can make a decision for rare events detection.

In addition, it is promising to measure rare events based on message importance and then analyze the relationship among the big data.
{The emerging applications related to big data require new ways to deal with anomalous detection or probability events mining.
To this end, we summarize some challenges and perspectives associated with rare events processing, which can be future research directions}
for information measures as shown in Table \ref{table.future}.
\begin{table*}[htb]%[!t]
\renewcommand{\arraystretch}{1.6}
% if using array.sty, it might be a good idea to tweak the value of
% \extrarowheight as needed to properly center the text within the cells
\caption{Perspective applications and use cases of rare events processing}
\label{table.future}
\newcommand{\tabincell}[2]{\begin{tabular}{@{}#1@{}}#2\end{tabular}}
\centering
%%%%%%%%%\resizebox{\textwidth}{40mm}{
% Some packages, such as MDW tools, offer better commands for making tables
% than the plain LaTeX2e tabular which is used here.
\begin{tabular}{|l|l|l|}%{|p{2.5cm}|p{1cm}|p{4.5cm}|}%{|c|c|c|}
\bottomrule
\rowcolor{mygray}
\textbf{Work Area} & \textbf{Related Work} & \tabincell{c}{\textbf{Key Points}}\\
\hline
Smart cities &
\tabincell{l} {
\cite{ref_A-Survey-on-data},
\cite{ref_An-architecture-for-the},
\cite{ref_An-anomaly-detection-in},
\cite{ref_IoT-driven-automated-object},
\cite{ref_Detecting-Urban-Anomalies-Using},\\
\cite{ref_Spatio-temporal-Anomaly},
\cite{ref_Uapd}
\cite{ref_Detecting-urban-black-holes},
\cite{ref_Scalable-graph-clustering-using},
\cite{ref_Graph-clustering-based-on},\\
\cite{ref_On-modularity-clustering},
\cite{ref_Spotting-significant-changing-subgraphs},
\cite{ref_Constraint-based-pattern-mining},
\cite{ref_Effective-density-queries-on},
\cite{ref_Mining-relaxed-temporal-moving},\\
\cite{ref_On-discovery-of-gathering},
\cite{ref_Identifying-attributing-and-describing},
\cite{ref_On-the-spatiotemporal-burstiness},
\cite{ref_A-context-aware-collaborative},
}
&
\tabincell{l} {
$\bullet$ Anomaly detection with urban surveillance data:\\
\ \ \ $\circ$  Security problems in wireless sensor networks \\
\ \ \ $\circ$ Optimizing transportation schedule system\\
\ \ \ $\circ$ The prediction problem in cities\\
$\bullet$ Urban Black Holes detection:\\
\ \ \ $\circ$ Detecting the groups of objects described by graphs\\
\ \ \ $\circ$ Multiple datasets analysis\\

}
\\
\hline
Automatic driving &
\tabincell{l} {
\cite{ref_Real-time-obstacles-detection},
\cite{ref_3D-Geometry-from-Planar},
\cite{ref_Real-Time-Obstacle-Detection},
\cite{ref_Obstacle-Detection-Using-Sparse},
\cite{ref_Obstacle-detection-based-on},\\
\cite{ref_Lost-and-found},
\cite{ref_A-real-time-low},
\cite{ref_A-head-wearable-short},
\cite{ref_Model-based-vehicle-detection},
\cite{ref_Object-detection-and-tracking},\\
\cite{ref_RefineNet},
\cite{ref_Selective-attention-for-detection},
\cite{ref_Simultaneous-localization-and-mapbuilding},
\cite{ref_Image-based-automatic-road},
\cite{ref_Automatic-crack-detection-on},\\
\cite{ref_Virtuous-visionbased-road-transportation},
\cite{ref_A-color-vision-based-lane},
\cite{ref_Road-curb-and-lanes-detection}
\cite{ref_Learning-Probabilistic},
\cite{ref_A-PSO-and-BFO-based},\\
\cite{ref_3D-Object-Proposals}
}
&
\tabincell{l} {
$\bullet$ Obstacles detection\\
$\bullet$ Detecting and tracking objects\\
$\bullet$ Detecting road surfaces and lanes\\
$\bullet$ probabilistic approaches and learning strategies
}
\\
\hline
\tabincell{l}{Applications to detection\\ in IoTs} &
\tabincell{l} {
\cite{ref_Recursive-principal-component-analysis},
\cite{ref_Outlier-Detection-Techniques-for},
\cite{ref_Research-Directions-for-the},
\cite{ref_A-lightweight-anomaly-mining},
\cite{ref_Detection-of-anomalies-in},\\
\cite{ref_Anomaly-detection-and-monitoring},
\cite{ref_Clustering-for-road-damage},
\cite{ref_Urban-anomaly-detection},
\cite{ref_Urban-sensing-and-smart},
\cite{ref_Performance-analysis-of-anomaly},\\
\cite{ref_Towards-an-emulated-IoT},
\cite{ref_Anomaly-Detection-and-Attribution},
\cite{ref_Outlier-Dirichlet-Mixture-Mechanism},
\cite{ref_An-Intelligent-Outlier-Detection-Method-With}
}
&
\tabincell{l} {
$\bullet$ Differentiating normal and abnormal data by use of\\ judging criterion\\
$\bullet$ Machine learning algorithms such as classification,\\ clustering and R-PCA
}
\\
\toprule
\end{tabular}%}
\end{table*}
\subsection{Smart Cities}
\subsubsection{Anomaly Detection for Urban Monitoring Data}
As a typical application of big data, smart city has been evolving rapidly with the increase of urban population.
This implies that cities can be monitored by countless devices in many aspects such as road traffic, transportation management, environment monitoring, healthcare, etc.
Actually, in cities, it is significant to detect the anomalies with small probability, which may provide effective guidance or warning information.

In order to investigate the anomaly detection problem in smart cities \cite{ref_A-Survey-on-data}, the major challenges are listed as follows.
\begin{itemize}
\item Security problems in the urban monitoring systems with wireless sensor networks (WSN).
\item The way to optimize the validity and reliability of transportation schedule system by avoiding the anomalies.
\item The long time prediction for the regular pattern of cities.
\item To distinguish the unexpected events from popular anomalies.
\item Automatic anomaly detection algorithms for the urban monitoring systems with IoTs.
\end{itemize}

In fact, the anomaly detection (or rare events detection) can be processed in many ways including machine learning, signal analysis and even information theory.
To be honest, there are some specific methods to detect the anomalies in smart cities, which may overcome the above challenges from different perspectives. Particularly,
%%%%==============================================================================================
in order to improve the security of the WSNs in urban monitoring systems,
a non-intrusive architecture is proposed to detect attacks by use of the support vector machine (SVM) \cite{ref_An-architecture-for-the}.
%%===============================================================
Moreover, for the IoTs of smart cities, by using automatic clustering or classification, the events with low probability can be identified in many applications such as the car parking scenario, polluted region monitoring and actionable bumps detection \cite{ref_An-anomaly-detection-in}.
In a wide sense, automatic target detection in urban surveillance systems can be also regarded as an anomaly detection \cite{ref_IoT-driven-automated-object}. As an example, there is a method presented to extract the regions with the highest energy frequency in pending images, which can help to reduce the complexity of detection.
{In addition, spatio-temporal data mining is also considered in urban anomaly detection.
Specifically, a two-step method (to compute individual anomaly scores (CIAS) and  to aggregate the individual anomaly scores (AIAS)) is proposed to give an anomaly score for each data source of each region at each time slot}
\cite{ref_Detecting-Urban-Anomalies-Using};
%H. Zhang, Y. Zheng, and Y. Yu.
%``Detecting Urban Anomalies Using Multiple Spatio-Temporal Data Sources,''
%\textit{Proceedings of the ACM on Interactive, Mobile,Wearable and Ubiquitous Technologies},vol. 2, no. 1, pp. 54:1--18, Mar. 2018.
{An improved Local Outlier Factor (LOF) algorithm (based on spatial-temporal cube) is adopted for abnormal region detection} \cite{ref_Spatio-temporal-Anomaly};
%Q. Wang, W. Lv, and B. Du,
%``Spatio-temporal Anomaly Detection in Traffic Data,''
%in \textit{Proceedings of the 2nd International Symposium on Computer Science and Intelligent Control (ISCSIC '18)}, Stockholm, Sweden, Sep. 2018, pp. 1--5.
{A Urban Anomaly PreDiction (UAPD) framework is designed to detect the anomalous change points and dig out the time-evolving inherent factors} \cite{ref_Uapd}.
%X. Wu, Y. Dong, C. Huang, J. Xu, D. Wang, and N. V. Chawla,
%``Uapd: Predicting urban anomalies from spatial-temporal data,''
%in \textit{Joint European Conference on Machine Learning and Knowledge Discovery in Databases}, Springer, Cham., Sep. 2017, pp. 622--638,

Notably, it is promising to exploit information theory to deal with anomaly detection by emphasizing the importance of rare events. By combining machine learning techniques, the importance measures focusing on rare events may provide new ways to cope with the anomaly detection and the evaluation of post processing, which plays an vital role in smart cities.

\subsubsection{Detecting Urban Black Holes}
As an important part of smart cities, the urban black hole denotes a region in which the whole traffic inflow is larger than the whole traffic outflow. Actually, the urban black hole can reflect emergencies or irregular events, namely rare events, including disasters, accidents, as well as traffic jams or congestion \cite{ref_Detecting-urban-black-holes,ref_A-context-aware-collaborative}.
It is worth detecting urban black holes efficiently, which can make a beneficial effect on urban safety. Therefore, some approaches are investigated as follows.
\begin{itemize}
\item Graph Clustering: With regard to the graph clustering, the approaches with the pruning schemes and the random matrix are proposed to characterize the potential black holes in a directed graph \cite{ref_Scalable-graph-clustering-using}. Besides, there are some other approaches detecting black holes by means of different measures \cite{ref_Graph-clustering-based-on,ref_On-modularity-clustering} such as attribute, modularity and density.
\item Dynamic Graph Detection: To detect black holes emerging in dynamic graph, some efficient approaches are proposed by means of the increment, pattern trees, and the pattern recognition with constraints \cite{ref_Spotting-significant-changing-subgraphs,ref_Constraint-based-pattern-mining}.
\item Groups Moving Recognition: On one hand, the density of regions is used to discover the object groups beyond the threshold during the observation time \cite{ref_Effective-density-queries-on}.
    On the other hand, moving together behaviors during a given time period are investigated to find out the tracking of a group of objects \cite{ref_Mining-relaxed-temporal-moving,ref_On-discovery-of-gathering}.
\item Spatio-Temporal Graph: Based on the spatio-temporal graph, some approaches are presented to mine spatial urban black holes \cite{ref_Identifying-attributing-and-describing}, as well as, detect the tracking of data temporally and spatially \cite{ref_On-the-spatiotemporal-burstiness}.
\end{itemize}

\begin{figure}[htb]%[!hbpt]%[!t]
\centering
\includegraphics[width=5.3in]{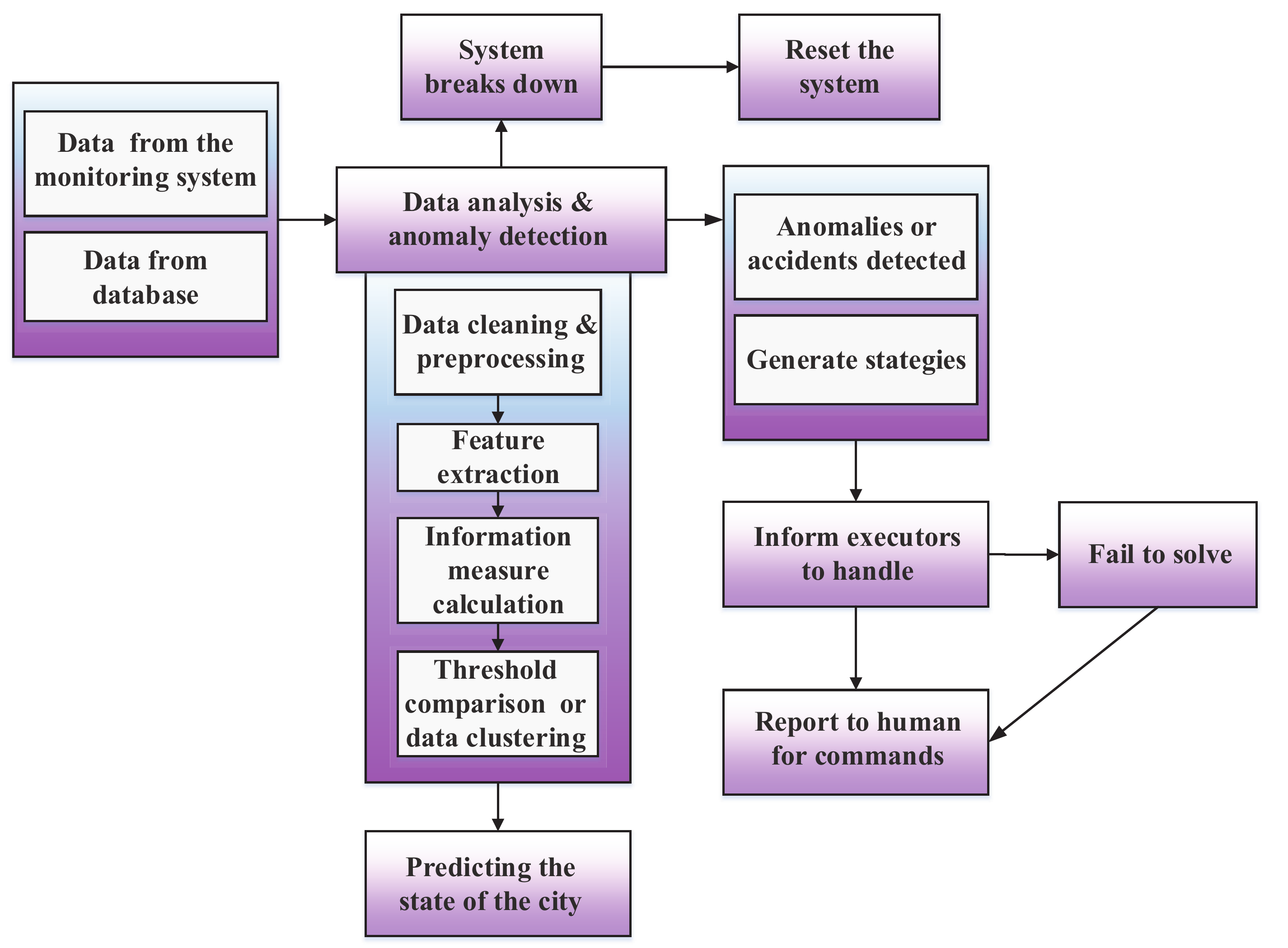}
\caption{
The rare events detection scheme for the smart city.
}
\label{fig_smart_city_chart}
\end{figure}

Actually,
from the perspective of probability distribution, it is possible to use information measures to find out urban black holes which may be described by graph methods.
To do so, a detection scheme for the smart city is shown in Fig. \ref{fig_smart_city_chart} whose details are as follows.

Specifically, data from the monitoring system and database are used to
detect the emergency events or accidents which can be regarded as the rare events.
Next, we can apply the information measures to analyze the relationships among events.
By using the data analysis and processing, the control center predicts the state of the city.
This can be adopted as a reference to update the system model and database.
Moreover, when anomalies are detected or the system breaks down, the control center can reset the system.
If anomalies or accidents are detected or some unsolved emergencies are reported, control center will take measures to handle them.
Then, if the system generates strategies for the anomaly detection, it would send commands to executors to solve the problems.
Besides, human can also set in the work directly when the system fails to finish the work.

\subsection{ Autonomous Driving }
As an important part of the autonomous driving,
obstacles detection makes a great influence on warning and predicting collisions and accidents
\cite{ref_Real-time-obstacles-detection}.
However, it is still a challenge to accurately detect the obstacles or objects with small probability in the view of computer vision. In general, some key issues of autonomous driving are summarized as follows.
\begin{itemize}
\item Obstacles Detection:
On one hand, some approaches are presented to characterize obstacles by use of image data \cite{ref_3D-Geometry-from-Planar},
v-disparity histogram \cite{ref_Real-Time-Obstacle-Detection},
as well as the models for the height-over ground \cite{ref_Obstacle-Detection-Using-Sparse,
ref_Obstacle-detection-based-on,ref_Lost-and-found}.
On the other hand,
deep learning tasks are used to detect obstacles by means of the image features and related information.
Moreover, a technique ``6D-vision'' is
also put forward to discover the dangerous events on the roads \cite{ref_A-real-time-low,ref_A-head-wearable-short}.
%%===============================================

\item Object Detection:
There are some approaches to detect and track objects by means of classification or clustering
\cite{ref_Model-based-vehicle-detection}.
The strategies and frameworks for object localization or tracking are also proposed depending on the Kalman filter \cite{ref_Object-detection-and-tracking}
and deep convolutional neural networks \cite{ref_RefineNet}.
Furthermore,
some other approaches are designed by use of the trade-off between
%saccades frequency and information loss,
camera orientations prediction and monitoring techniques
\cite{ref_Selective-attention-for-detection,
ref_Simultaneous-localization-and-mapbuilding}.

\item Detecting Road Surface and Lanes:
As for road surface detection, the discriminant analysis (DA) is presented to characterize the road crack \cite{ref_Image-based-automatic-road,ref_Automatic-crack-detection-on}. This can provide a threshold for classification according to the road texture and color in images.
%Besides, cracks can also be detected by calculating the pixel variance in road pictures.
%Additionally,
Besides, in order to detect the road curb and lanes, it is common to regard
color and texture as interesting features of roads.
These can be used by combining classification with the hue-saturation-intensity (HSI) color space or red-green-blue (RGB) color space
\cite{ref_Virtuous-visionbased-road-transportation,ref_A-color-vision-based-lane}. Besides, another framework of road curb and lanes detection is addressed by extracting the 3D parameters from some curb models \cite{ref_Road-curb-and-lanes-detection}.
\end{itemize}
{Moreover, there are some works proposed based on probabilistic approaches and learning strategies.
Gaussian process (GP) regression decomposition based on a superpixel-like algorithm is employed to validate quasi-constant velocity models which build a set of Kalman filters to identify the abnormal motions online} \cite{ref_Learning-Probabilistic}.
{A particle swarm optimization (PSO) and bacterial foraging optimization (BFO)-based learning strategy (PBLS) is presented to improve the classifier and loss function of strengthened region proposal network (SRPN), which can be applied in object detection of autonomous driving} \cite{ref_A-PSO-and-BFO-based}.
{A set of 3D object proposals based on an energy function are obtained to detect high-quality 3D objects by use of a convolutional neural net (CNN)} \cite{ref_3D-Object-Proposals}.

\begin{figure}[htb]%[!hbpt]%[!t]
\centering
\includegraphics[width=5.5in]{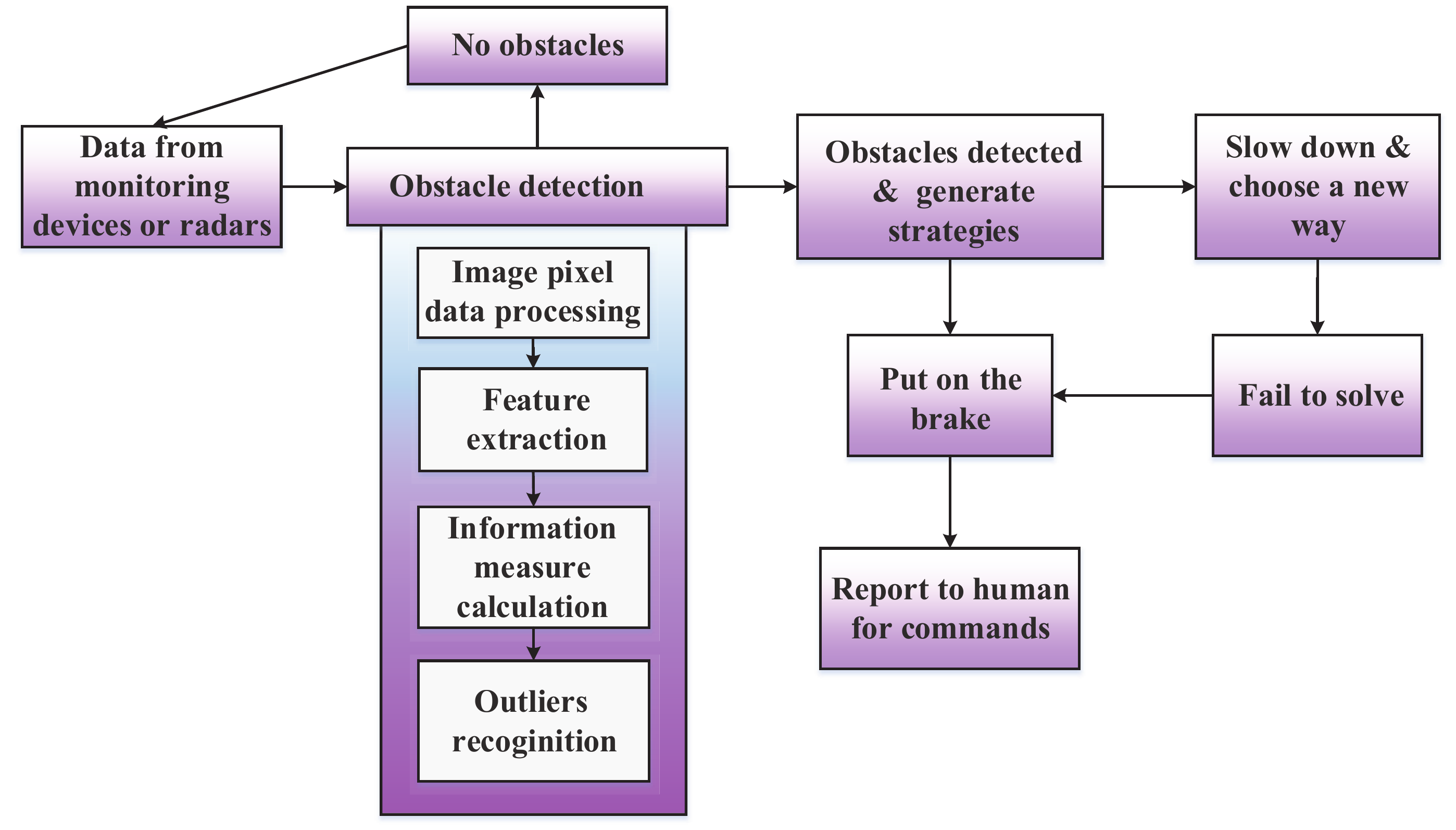}
\caption{
The obstacle detection scheme for the autonomous driving.
}
\label{fig_driving_chart}
\end{figure}

Additionally, with regard to the detection for autonomous driving, it is apparent that rare events play important roles in many aspects of vehicular safety system.
%Besides, small probability events can be characterized by some message importance measures.
By measuring small probability, it is appropriate to apply information theory to the autonomous driving detection.
To do so, an obstacle detection scheme is shown in Fig. \ref{fig_driving_chart}, whose details are given as follows.
Based on the data from monitoring devices or radars, the autonomous control system can detect obstacles or other outlier events, which can be analyzed by use of information measures.
If no obstacle is discovered, the system will continue the normal surveillance.
However, if some obstacles are detected, the system will take measures to solve the problem by slowing down and choosing a new way.
When the emergencies are not solved well, it will put on the brake and report them to drivers for further commands.

\subsection{Applications on Detection in IoTs}
Outlier detection in IoTs, is to dig out the minority of sensors data exactly \cite{ref_Recursive-principal-component-analysis,ref_Outlier-Detection-Techniques-for}. In fact, it is essential to differentiate the outlier data or observations
from the normal data so that one can gain the warning information and prevent the outlier data from misleading us \cite{ref_Research-Directions-for-the}. There exist a various of researches focusing on the outlier detection which are also considered to detect rare events in IoTs systems.

On one hand, there are some approaches to detect outliers in IoTs directly, such as using
%%=======================================================
%A Lightweight Anomaly Mining Algorithm in the Internet of Things
Jaccard coefficient or Euclidean distance as the criterion of decision making \cite{ref_A-lightweight-anomaly-mining},
%Detection of Anomalies in Data for Monitoring of Security Components in the Internet of Things
referring to the expert knowledge on security \cite{ref_Detection-of-anomalies-in}, as well as,
%Anomaly Detection and Monitoring in Internet of Things Communication
monitoring the abnormal traffic among communication devices \cite{ref_Anomaly-detection-and-monitoring}, etc.
%%========================
On the other hand,
several researches divide observations into different groups to find out the outliers by use of classification and clustering algorithms \cite{ref_Clustering-for-road-damage,ref_Urban-anomaly-detection}.
To address this kind of matter, a few approaches also introduce static data series \cite{ref_Urban-sensing-and-smart} or dynamic time series into the machine learning algorithms.
Besides, a framework of data analysis is put forward by means of the recursive principal component analysis (R-PCA) \cite{ref_Recursive-principal-component-analysis}, which provides another way to investigate the security of IoTs systems.

In light of the fact that the data observed from IoTs are usually fed to cloud service systems, some approaches are proposed by
blending both IoTs and cloud technologies \cite{ref_Performance-analysis-of-anomaly}.
Moreover, to test IoTs systems conveniently, a new method is presented to emulate the environments of IoTs by means of a network emulator, which can improve the processing efficiency for outlier detection \cite{ref_Towards-an-emulated-IoT}.

{Furthermore, some probabilistic models and large-scale processing approaches are also exploited in the anomalies detection of IoTs.
A statistical decision framework based on temporally correlated traffic is designed, which develops two low-complexity algorithms (based on cross entropy method and generalized likelihood ratio test) to achieve anomaly detection and attribution} \cite{ref_Anomaly-Detection-and-Attribution}.
{An adversarial statistical learning mechanism, outlier Dirichlet mixture-based anomaly detection systems (ODM-ADS), is presented to obtain legitimate profiles and discover suspicious anomalies} \cite{ref_Outlier-Dirichlet-Mixture-Mechanism}.
{Besides, there are two methods are proposed, namely a one-class support Tucker machine (OCSTuM) and an OCSTuM based on a genetic algorithm called GA-OCSTuM, which extend one-class support vector machines to tensor space to detect anomalies in IoTs} \cite{ref_An-Intelligent-Outlier-Detection-Method-With}.

However, in spite of many efficient approaches for outlier detection, few researches consider to exploit the small probability character in the viewpoint of probability distribution.
Actually, it is promising to take use of information measures to analyze the outliers of IoTs.

\begin{figure}[htb]%[!hbpt]
\centering
\includegraphics[width=5.5in]{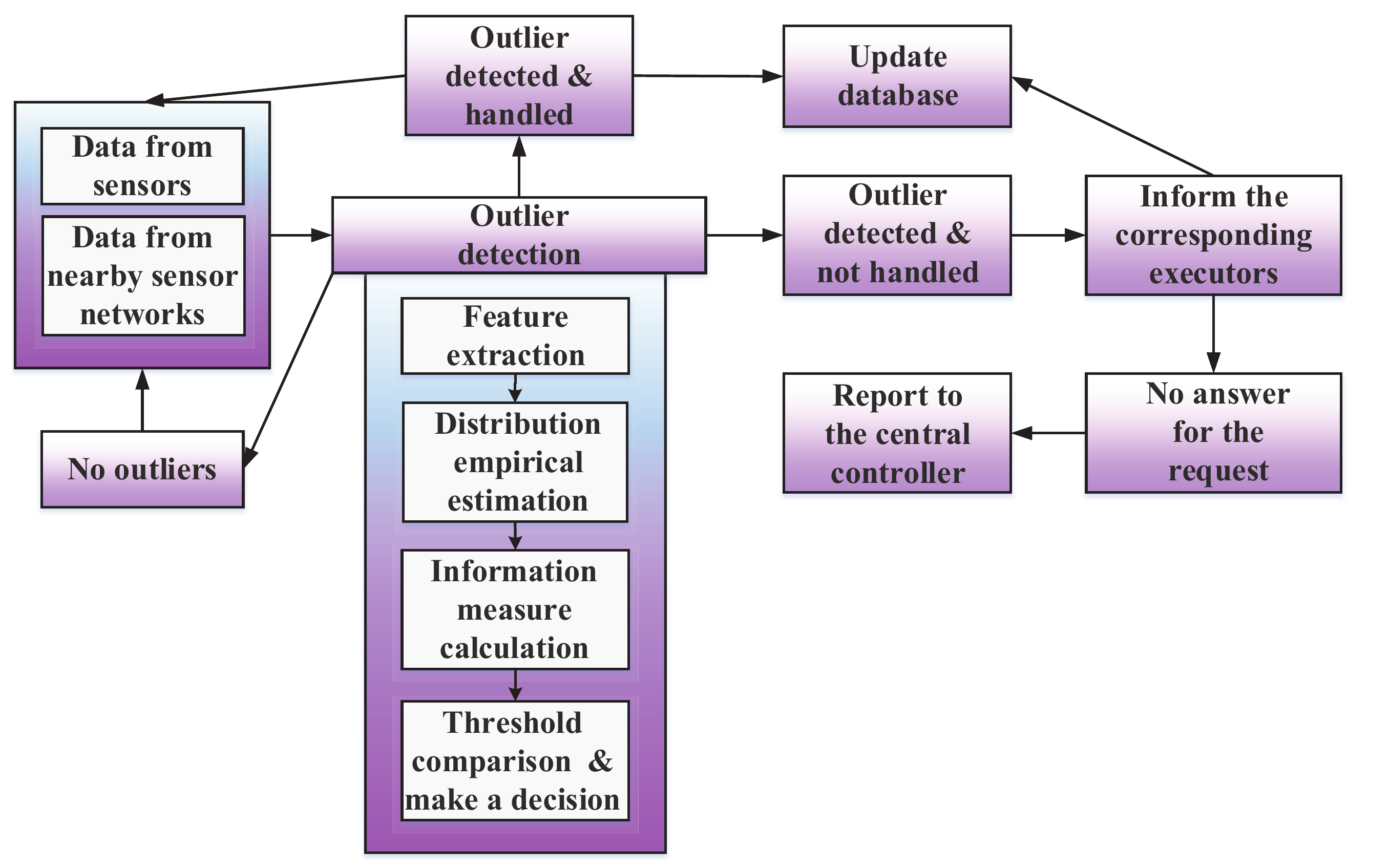}
\caption{
The outlier detection scheme for the IoTs composed of monitoring sensors.
}
\label{fig_IoT_chart}
\end{figure}

From the perspective of information theory, importance measures can provide a specific access to tackle the outlier detection problem by using probability distribution, which is shown in Fig. \ref{fig_IoT_chart} whose details are as follows.
The data collected by distributed sensors are used to detect the potential or ongoing outliers by resorting to information measures.
If an outlier is detected and handled, the local center will continue to collect data and update the database.
However, if a detected outlier is not handled well, the local center will contact with executors to solve the problem and save the data to the database. Once there is no answer
for the request, local center will report it to the control center.

\section{Conclusion}\label{section6}
In this paper, we gave a total review on information measures for rare events in big data. % from the perspective of information theory.
In order to characterize the importance of rare events, we summarized some message  measures such as the parametric MIM and the non-parametric MIM
which have properties on emphasizing small probability elements for a given distribution.
These information measures are regarded as promising criterions or tools for statistical big data analytics.
Furthermore, we introduced that measures focusing on rare events can provide new ways for message processing such as compression and transmission.
Moreover, some other applications in big data have been discussed including efficient estimation, dimension reduction and relationship analysis.
{Additionally, we introduced that information measures for rare events could be applicable for some future research directions including smart cities, autonomous driving, and anomaly detection in IoTs.
In these cases, there exist several future challenges of information measures summarized as follows:

\textit{i)} Data storage and low latency computation for the data sets containing rare events.

\textit{ii)} Feature extraction and data cleaning of holding rare events.

\textit{iii)} Design of information theoretic criterions to measure distributions while considering the values of rare events.

\textit{iv)} Efficient methods of information measure estimations.

\textit{v)} Correlation and causality analysis based on information measures.

\textit{vi)} Decision making strategies for rare events (or probability events) mining.
}
%\appendices
%% you can choose not to have a title for an appendix
%% if you want by leaving the argument blank
%\section{}

\section*{Acknowledgment}
The authors indeed appreciate the support of the China Major State Basic Research Development Program (973 Program) No.2012CB316100(2), and the National Natural Science Foundation of China (NSFC) No. 61771283.

\ifCLASSOPTIONcaptionsoff
  \newpage
\fi

%% biography section
%%
%% If you have an EPS/PDF photo (graphicx package needed) extra braces are
%% needed around the contents of the optional argument to biography to prevent
%% the LaTeX parser from getting confused when it sees the complicated
%% \includegraphics command within an optional argument. (You could create
%% your own custom macro containing the \includegraphics command to make things
%% simpler here.)
%%\begin{IEEEbiography}[{\includegraphics[width=1in,height=1.25in,clip,keepaspectratio]{mshell}}]{Michael Shell}
%% or if you just want to reserve a space for a photo:
%\end{IEEEbiography}
%% insert where needed to balance the two columns on the last page with
%%\newpage
%% You can push biographies down or up by placing
%% a \vfill before or after them. The appropriate
%% use of \vfill depends on what kind of text is
%% on the last page and whether or not the columns
%% are being equalized.
%
%%\vfill
%
%% Can be used to pull up biographies so that the bottom of the last one
%% is flush with the other column.
%%\enlargethispage{-5in}


\begin{thebibliography}{00}



\bibitem{ref_Big-IoT-data-analytics}
M. Marjani, F. Nasaruddin, A. Gani, A. Karim, I. Hashem, A. Siddiqa, and I. Yaqoob,
``Big IoT data analytics: Architecture, opportunities, and open research challenges,''
\textit{IEEE Access}, {vol. 5}, pp. 5247--5261, Mar. {2017}.

\bibitem{ref_Big-Data-Deep-Learning}
X. Chen and X. Lin,
``Big data deep learning: challenges and perspectives,''
\textit{IEEE Access}, {vol. 2}, pp. 514--525, May. {2014}.

\bibitem{ref_Toward-scalable-systems-for-big-data-analytics}
H. Hu, Y. Wen, T. Chua, and X. Li,
``Toward scalable systems for big data analytics: A technology tutorial,''
\textit{IEEE Access}, {vol. 5}, pp. 7776--7797, Jun. {2017}.

\bibitem{ref_Machine-learning-with-big-data}
A. Leureux, K. Grolinger, H. Elyamany, and M. Capertz,
``Machine learning with big data: Challenges and Approahces,''
 \textit{IEEE Access}, {vol. 5}, pp. 2169--3536, Apr. {2017}.

\bibitem{ref_Information-security-in-big-data}
L. Xu, C. Jiang, J. Wang, J. Yuan, and Y. Ren,
``Information security in big data: Privacy and data mining,''
\textit{IEEE Access}, {vol. 2}, pp. 1149--1176, Oct. {2014}.

%%%%==============================
\bibitem{ref_Deep-learning}
Y. LeCun, Y. Bengio, and G. Hinton,
Deep learning,
\textit{Nature}, {vol. 521}, pp. 436--444, May. {2015}.

\bibitem{ref_DNN-filter-bank}
H. Yu, Z. Tan, Y. Zhang, Z. Ma, and J. Guo,
``DNN filter bank cepstral coefficients for spoofing detection,''
\textit{IEEE Access}, {vol. 5}, pp. 4779--4787, Mar. {2017}.

\bibitem{ref_Text-independent-speaker}
Z. Ma, H. Yu, Z. Tan, and J. Guo,
``Text-independent speaker identification using the histogram transform model,''
\textit{IEEE Access}, {vol. 4}, pp. 9733--9739, {2016}.
%%%%%======================================
\bibitem{Lu_Fan}
Y. Lu, K. Xiong, P. Fan, Z. Zhong, and K. Letaief,
``Robust Transmit Beamforming With Artificial Redundant Signals for Secure SWIPT System Under Non-Linear EH Model,''
\textit{IEEE Trans. Wirel. Commun.}, {vol. 17}, no. 4, pp. 2218--2232, Jan. {2018}.

\bibitem{X_Fan}
K. Xiong, C. Chen, G. Qu, P. Fan, and K. Letaief,
``Group cooperation with optimal resource allocation in wireless powered communication networks,''
\textit{IEEE Trans. Wirel. Commun.}, {vol. 16}, no. 6, pp. 3840--3853, Mar. {2017}.

\bibitem{X2_Fan}
K. Xiong, P.Fan, Y. Lu, and K. Letaief,
``Energy Efficiency with Proportional Rate Fairness in Multi-Relay OFDM
Networks,''
\textit{IEEE J. Sel. Areas Commun.}, {vol. 34}, no. 5, pp. 1431--1447, May. {2016}.

\bibitem{X3_Fan}
Y. Lu, K. Xiong, P. Fan, Z. Ding, Z. Zhong, and K. Letaief,
``Global energy efficiency in secure MISO SWIPT systems with non-linear power-splitting EH model,''
\textit{IEEE J. Sel. Areas Commun.}, {vol. 37}, no. 1, pp. 216--232, Jan. {2019}.
%%======================

\bibitem{ref_fingerprinting-fusion}
B. Koo, S. Lee, M. Lee, D. Lee, S. Lee, and S. Kim,
``PDR{\//}fingerprinting fusion indoor location tracking using RSS recovery and clustering,''
in \textit{Proc. International Conference on Indoor Positioning and Indoor Navigation (IPIN)}, Busan, South Korea, Sep. 2015, pp. 699--704.

%%==================================
\bibitem{ref_Variational-bayesian-matrix}
Z. Ma, A. Teschendorff, A. Leijon, Y. Qiao, H. Zhang, and J. Guo,
``Variational bayesian matrix factorization for bounded support data,''
\textit{IEEE Trans. Pattern Anal. Mach. Intell.}, {vol. 37}, no. 4, pp. 876--889, Sep. {2014}.

\bibitem{ref_Decorrelation-of-neutral}
Z. Ma, J. Xue, A. Leijon, Z. Tan, Z. Yang, and J. Guo,
``Decorrelation of neutral vector variables: theory and applications,''
\textit{IEEE Trans. Neural Netw. Learn. Syst.}, {vol. 29}, no. 1, pp. 129--143, Jan. {2018}.

\bibitem{ref_Vector-quantization}
Z. Ma, A. Leijon, and W. Kleijn,
``Vector quantization of LSF parameters with a mixture of dirichlet distributions,''
\textit{IEEE Trans. Audio, Speech, Language Process}, {vol. 21}, no. 9, pp. 1777--1790, Jan. {2013}.

\bibitem{ref_Network-denoising}
H. Gao, X. Wang, J. Tang, and H. Liu,
``Network denoising in social media,''
in \textit{Proc. 2013 IEEE/ACM International Conference on Advances in Social Networks Analysis and Mining (ASONAM 2013)}, Niagara Falls, Canada, Apr. 2014, pp. 564--571.

\bibitem{ref_Making-better-use}
Y. Qi, Y. Song, and T. Xiang,
``Making better use of edges via perceptual grouping,''
in \textit{Proc. 2015 IEEE Conference on Computer Vision and Pattern Recognition (CVPR)}, Boston, USA, Oct. 2015, pp. 1856--1865.


\bibitem{ref_Data-intensive-applications}
C. Chen and C. Zhang,
``Data-intensive applications, challenges, techniques and technologies: A survey on Big Data,''
\textit{Inf. Sci.}, {vol. 275}, pp. 314--347, {2014}.

\bibitem{ref_Preconditioned-data-sparsification}
F. Pourkamali-Anaraki and S. Becker,
``Preconditioned data sparsification for big data with applications to pca and k-means,''
\textit{IEEE Trans. Inf. Theory}, {vol. 63}, no. 5, pp. 2954--2974, Feb, {2017}.


\bibitem{ref_A-comprehensive-survey-of-data-mining-based-accounting-fraud-detection-research}%3
S. Wang,
``A comprehensive survey of data mining-based accounting-fraud detection research,''
in \textit{Proc. IEEE Intelligent Computation Technology and Automation (ICICTA)}, Madurai, India, May. 2010, pp. 50--53.

\bibitem{ref_Fraudulent-Financial-Reporting}
``Fraudulent Financial Reporting: Consideration of Industry Traits and Corporate Governance Mechanisms,''
\textit{Accounting Horizons}, vol. 14, no. 4, pp. 441--454, Dec. 2000.
%%=================================================

\bibitem{ref_Data-Mining-Approaches-for-Intrusion-Detection}%1
W. Lee and S. Stolfo,
``Data Mining Approaches for Intrusion Detection,''
in \textit{Proc. Usenix security}, San Antonio, USA, Jan. 1998, pp. 291--300.

\bibitem{ref_Mining-intrusion-detection-alarms-for-actionable-knowledge}%2
K. Julisch and M. Dacier,
``Mining intrusion detection alarms for actionable knowledge,''
in \textit{Proc. Acm International Conference on Knowledge Discovery \& Data Mining}, New York, USA, Aug. 2002, pp. 366-375.

%%-----------------------
\bibitem{ref_Research-on-intrusion-detection}
X. Zhang and X. Hao,
``Research on intrusion detection based on improved combination of K-means and multi-level SVM,''
in \textit{Proc. 2017 IEEE 17th International Conference on Communication Technology (ICCT)},
Chengdu, China, Oct. 2017, pp. 2042--2045.

\bibitem{ref_Application-of-CART-decision}
M. Li,
``Application of CART decision tree combined with PCA algorithm in intrusion detection,''
in \textit{Proc. 2017 8th IEEE International Conference on Software Engineering and Service Science (ICSESS)},
Beijing, China, Nov. 2017, pp. 38--41.

\bibitem{ref_Neural-network-based}
G. Karatas and O. Sahingoz,
``Neural network based intrusion detection systems with different training functions,''
in \textit{Proc. 2018 6th International Symposium on Digital Forensic and Security (ISDFS)},
Antalya, Turkey, Mar. 2018, pp. 1--6.

%%=============================================================================




%%---------------------

%%============================================================================================

\bibitem{ref_Event-analysis-for-security-incident-management}
V. Desnitsky and I. Kotenko,
``Event analysis for security incident management on a perimeter access control system,''
in \textit{Proc. 2016 XIX IEEE International Conference on Soft Computing and Measurements (SCM)}, St. Petersburg, Russia, May. 2016, pp. 481--483.

\bibitem{ref_Securing-embedded-systems}
D. Hwang, P. Schaumont, K. Tiri, and I. Verbauwhede,
``Securing embedded systems,''
\textit{IEEE Security and Privacy}, {vol. 4}, pp. 40--49, Apr. {2006}.

\bibitem{ref_Security-in-embedded-systems}
S. Ravi, A. Kocher, and S.Hattangady,
``Security in embedded systems: design challenges,''
\textit{ACM Transactions on Embedded Computing Systems}, vol, 3, no. 3, pp. 461--491, Aug. {2014}.

\bibitem{ref_Blockchains-and-Smart-Contracts}
K. Christidis and M. Devetsikiotis,
''Blockchains and Smart Contracts for the Internet of Things,''
\textit{IEEE Access}, vol. 4, pp. 2292--2303, June. 2016.

\bibitem{ref_Design-and-realization}
J. Wu and W. Zhao, ''Design and realization of winternet: From net of things to internet of things,''
\textit{ACM Trans. Cyber Phys. Syst.}, vol. 1, no. 1, Feb. 2017, Art. no. 2.

\bibitem{ref_A-Survey-on-Internet}
J. Lin, W. Yu, N. Zhang, X. Yang, H. Zhang, and W. Zhao, ''A Survey on Internet of Things: Architecture, Enabling Technologies, Security and Privacy, and Applications,''
\textit{IEEE Internet Things J.}, vol. 4, no. 5, pp. 1125--1142, Oct. 2017.

\bibitem{ref_Internet-of-things-and-big-data}
Y. Sun, H. Song, A. J. Jara, and R. Bie, ''Internet of things and big data analytics for smart and connected communities,''
\textit{IEEE Access}, vol. 4, pp. 766--773, Mar. 2016.

%%==============================================================
\bibitem{ref_SAlightweight-anomaly-detection-technique}
H. Sedjelmaci, S.M.Senouci, and M. Al-Bahri, ''Alightweight anomaly detection technique for low-resource iot devices: A game-theoretic methodology,'' in \textit{Proc. 2016 IEEE International Conference on Communications (ICC)}, Kuala Lumpur, Malaysia, May. 2016, pp. 1--6.

\bibitem{ref_Performance-and-accuracy-trade-off-analysis}
P. Souza, W. Marques, F. Rossi, G. Rodrigues, and R. Calheiros, ''Performance and accuracy trade-off analysis of techniques for anomaly detection in IoT sensors,'' in \textit{Proc. 2017 International Conference on Information Networking (ICOIN)}, Da Nang, Vietnam, Jan. 2017, pp. 486--491.

\bibitem{ref_SVELTE}
S. Raza, L. Wallgren, and T. Voigt, ``SVELTE: Real-time intrusion detection in the Internet of Things'', \textit{Ad Hoc Networks}, vol. 11, no. 8, pp. 2661--2674, Nov. 2013.

\bibitem{ref_A-lightweight-anomaly-detection-technique}
H. Sedjelmaci, S.M.Senouci, and M. Al-Bahri, ''A lightweight anomaly detection technique for low-resource IoT devices: A game-theoretic methodology,'' in \textit{Proc. 2016 IEEE International Conference on Communications (ICC)}, Kuala Lumpur, Malaysia, May. 2016, pp. 1--6.

\bibitem{ref_Anomaly-detection-in-environmental}
J. C. Bezdek, S. Rajasegarar, M. Moshtaghi, C. Leckie,  M. Palaniswami, and T. C. Havens  ``Anomaly detection in environmental monitoring networks,'' \textit{ IEEE Comp. Int. Mng.}, vol. 6, No. 2, pp. 51--58, Apri. 2011.

\bibitem{ref_Internet-of-Things-for-Smart-Cities}
A. Zanella, N. Bui, and M. Zorzi, ``Internet of Things for smart cities,'' \textit{IEEE Internet of Things Journal}, vol. 1, no. 1, pp. 22--32, Feb. 2014.

\bibitem{ref_An-anomaly-detection-in-smart-cities}
R. Jain and H. Shah, ''An anomaly detection in smart cities modeled as wireless sensor network,''
in \textit{Proc. 2016 International Conference on Signal and Information Processing (IConSIP)}, Nanded, India, Oct. 2016, pp. 1--5.

\bibitem{ref_Dynamic-network-model-for-smart-city}
O. Kotevska, A. G. Kusne, D. V. Samarov, A. Lbath, and A. Battou,
``Dynamic network model for smart city data-loss resilience case study: city-to-city network for crime analytics,'' \textit{IEEE Access}, vol. 5, pp. 20524--20535, Oct. 2017.

\bibitem{ref_Flexible-spectrum-management-in-a-smart-city}
E. Markova, I. Gudkova, A. Ometov, I. Dzantiev, and S. Andreev, Y. Koucheryavy, and K. Samouylov,
``Flexible spectrum management in a smart city within licensed shared access framework,'' \textit{IEEE Access}, vol. 5, pp. 22252--22261, Oct. 2017.
%%--------------------

\bibitem{ref_Smart-city}
S. Mallapuram, N. Ngwum, F. Yuan, C. Lu, and W. Yu,
``Smart city: The state of the art, datasets, and evaluation platforms,''
in \textit{Proc. 16th IEEE/ACIS Int. Conf. Comput. Inf. Sci. (ICIS)}, Wuhan, China, May 2017, pp. 447--452.

\bibitem{ref_To-Smart-City}
S. Wan, J. Lu, P. Fan, K. B. Letaief,
``To Smart City: Public Safety Network Design for Emergency,''
\textit{IEEE Access}, vol. 6, pp. 1451--1460 , Dec. 2017.
%%----------------------
\bibitem{ref_Efficient-algorithms-for-mining-outliers-from-large-data-sets}%4
S. Ramaswamy, R. Rastogi, and K. Shim. ''Efficient algorithms for mining outliers from large data sets,'' \textit{ACM SIGMOD Record}, vol. 29, no. 2, pp. 427-438, May. 2000.

%%----------------------
\bibitem{ref_Improved-principal-component}
F. Harrou, F. Kadri, S. Chaabane, C. Tahon, and Y. Sun, ''Improved principal component analysis for anomaly detection: Application to an emergency department,''
\textit{Comput. Ind. Eng.}, vol. 88, pp. 63--77, Oct. 2015.

\bibitem{ref_An-improved-methodology}
S. Xu, M. Baldea, T. F. Edgar, W. Wojsznis, T. Blevins, and M. Nixon, ''An improved methodology for outlier detection in dynamic datasets,''
\textit{AIChE J.}, vol. 61, no. 2, pp. 419--433, 2015.

\bibitem{ref_Nonlinear-Gaussian-belief}
H. Yu, F. Khan, and V. Garaniya, ''Nonlinear Gaussian belief network based fault diagnosis for industrial processes,''
\textit{J. Process Control}, vol.35, pp. 178--200, Nov. 2015.

\bibitem{ref_Principal-components-selection}
A. Prieto-Moreno, O. Llanes-Santiago, and E.Garc\'{i}a-Moreno, ``Principal components selection for dimensionality reduction using discriminant information applied to fault diagnosis,''
\textit{J. Process Control}, vol. 33, pp. 14--24, Sep. 2015.

%%=============
\bibitem{ref_Traffic-safety-facts}
NHTSA, ``Traffic safety facts 2011,'' NHTSA, Tech. Rep., 2011, Available :http://www-nrd.nhtsa.dot.gov/Pubs/811754AR.pdf.

\bibitem{ref_Detecting-unexpected-obstacles-for-self-driving-cars}
S. Ramos, S.Gehrig, P. Pinggera, U. Franke, and C. Rother, ''Detecting unexpected obstacles for self-driving cars: Fusing deep learning and geometric modeling,'' in \textit{Proc. 2017 IEEE Intelligent Vehicles Symposium (IV)}, Redondo Beach, USA, June. 2017, pp. 1025--1032.

\bibitem{ref_Lane-following-and-obstacle-detection-techniques}
P. Amaradi, N. Sriramoju, L. Dang, G. S. Tewolde, and Ja. Kwon,
''Lane following and obstacle detection techniques in autonomous driving vehicles,'' in
\textit{Proc. 2016 IEEE International Conference on Electro Information Technology (EIT)},
North Dakota, USA, May. 2016, pp. 0674--0679.

\bibitem{ref_An-improved-lane-departure-method}
V. Gaikwad and S. Lokhande, ``An improved lane departure method for advanced driver assistance system,'' in \textit{Proc. International Conference on Computing, Communication and Applications (ICCCA)}, Dindigul, India, Feb. 2012, pp. 1--5.

\bibitem{ref_Implementation-of-lane-detection-system}
S. Lee, H. Son, and K. Min, ''Implementation of lane detection system using optimized Hough transform circuit,'' in \textit{2010 IEEE Asia Pacific Conference on Circuits and Systems (APCCAS)}, Kuala Lumpur, Malaysia, Dec. 2010, pp. 406--409.

%%=================================
\bibitem{ref_Region-based-convolutional-networks}
R. Girshick, J. Donahue, T. Darrell, and J. Malik, ''Region-based convolutional networks for accurate object detection and semantic segmentation,''
\textit{IEEE Trans. Pattern Anal. Mach. Intell.}, vol. 38, no. 1, pp. 142--158, Jan. 2016.

\bibitem{ref_Object-detection-with}
P.F. Felzenszwalb, R. B. Girshick, D. McAllester and D. Ramanan, ``Object detection with discriminatively trained part-based models,''
\textit{IEEE Trans. Pattern Anal. Mach. Intell.}, vol. 32, no. 9, pp. 1627--1645, Spe. 2010.

%%===================================================================
\bibitem{ref_Traffic-flow-prediction}
Y. Lv, Y. Duan, W. Kang, Z. Li, and F. Wang, ''Traffic flow prediction with big data: a deep learning approach,''
\textit{IEEE Trans. Intell. Transp. Syst.}, vol. 16, no. 2, pp. 865--873, Apr. 2015.

%%=======================================

\bibitem{ref_Real-time-obstacles-detection-and-status-classification}
W. Song, Y. Yang, M. Fu, F. Qiu, and M. Wang,
''Real-time obstacles detection and status classification for collision warning in a vehicle active safety system,'' \textit{IEEE Trans. Intell. Transp. Syst.}, vol. PP, no. 99, pp. 1--16, 2017.

%%========================================================

\bibitem{ref_Anomaly-detection-A-survey}%5
V. Chandola, A. Banerjee, and V. Kumar. ''Anomaly detection: A survey,'' \textit{ACM computing surveys (CSUR)}, vol.41, no.3, pp. 1--58, July. 2009.

\bibitem{ref_Multiblock-independent-component}
Q. Jiang, B. Wang, and X. Yan, ''Multiblock independent component analysis integrated with Hellinger distance and Bayesian inference for non-Gaussian plant-wide process monitoring,''
\textit{Ind. Eng. Chem. Res.}, vol. 54, no. 9, pp. 2497--2508, 2015.

\bibitem{ref_Information-theoretic-measures-for-anomaly-detection}%6
W. Lee and D Xiang. ''Information-theoretic measures for anomaly detection,'' in \textit{Proc. IEEE Symposium on Security and Privacy 2001}, Oakland, USA, May. 2001, pp. 130-143.

\bibitem{ref_An-information-theoretic-approach-to-detection-of-minority-subsets-in-database}%7
S. Ando and E. Suzuki. ''An information theoretic approach to detection of minority subsets in database,'' in \textit{Proc. IEEE Sixth International Conference on Data Mining}, Hong Kong, China, Dec. 2006, pp. 11-20.

\bibitem{ref_Kullback-Leibler-Divergence-(KLD)-Based-Anomaly-Detection}%\label{ref7}
A. Anderson and H. Haas, ''Kullback-Leibler Divergence (KLD) based anomaly detection and monotonic sequence analysis,'' in \textit{Proc. IEEE Vehicular Technology Conference (VTC Fall) }, San Francisco, USA, Sep. 2011, pp. 1--5.

\bibitem{ref_The-large-deviation}
H. Touchette.
''The large deviation approach to statistical mechanics,''
\textit{Physics Reports}, vol. 478, no. 1--3, pp. 1--69, Jul. 2009.

\bibitem{ref_A-measure-of-the-concentration}
R. P. Curiel and S. Bishop.
''A measure of the concentration of rare events,''
\textit{Sci. Rep.}, vol. 6, no. 32369, pp. 1--6, Aug. 2016.

\bibitem{ref_A-large-deviations-approach}
N. Weinberger and N. Merhav,
''A large deviations approach to secure lossy compression,''
\textit{IEEE Trans. Inf. Theory}, vol. 63, no. 4, pp. 2533--2559, Apr. 2017.

\bibitem{ref_statistical-traffic-anomaly-detection}
I. C. Paschalidis and G. Smaragdakis,
``A large deviations approach to statistical traffic anomaly detection,''
in \textit{Proc. IEEE Conference on Decision \& Control (CDC)},
San Diego, USA, Dec. 2006, pp. 1900--1905.

\bibitem{ref_Analytical-model-of-the-KL-Divergence}%\label{ref2}
A. Youssef, C. Delpha, and D. Diallo, ``Analytical model of the KL Divergence for Gamma distributed data: application to fault estimation,'' in \textit{Proc. IEEE Signal Processing Conference (EUSIPCO)}, Nice, France, Aug. 2015, pp. 1--6.

\bibitem{ref_Key-frame-selection-based-on-KL-divergence}%\label{ref3}
L. Li, Q. Xu, X. Luo, and S. Sun, ''Key frame selection based on KL-divergence,'' in \textit{Proc. IEEE International Conference on Multimedia Big Data (BigMM)}, Beijing, China, Apr. 2015, pp. 1--6.

\bibitem{ref_Human-Ear-recognition-based-on-Multi-scale}%\label{ref5}
Z. Youbi, L. Boubchir, and M. D. Bounneche, et al., ''Human Ear recognition based on Multi-scale Local Binary Pattern descriptor and KL divergence,'' in \textit{Proc. IEEE Telecommunications and Signal Processing (TSP) }, Vienna, Austria, Jun. 2016, pp. 1--6.

\bibitem{ref_A-Study-on-Invariance-of-$f$-Divergence-and-Its-Application}%\label{ref6}
Y. Qiao and N. Minematsu, ''A Study on invariance of $f$-Divergence and its application to speech recognition,'' \textit{IEEE Trans. Signal Process.}, vol. 58, no. 7, pp. 3884--3890, July. 2010.

%%====================================================
\bibitem{ref_Amplifying-inter-message-distance}
R. She, S. Y. Liu, and P. Fan,
``Amplifying inter-message distance: on information divergence measures in big data,''
\textit{IEEE Access}, vol. 5, pp. 24105--24119, 2017.

\bibitem{ref_Linear-complexity-exponentially-consistent}
Y. Bu, S. Zou, and V. V. Veeravalli,
''Linear-complexity exponentially-consistent tests for universal outlying sequence detection,''
in \textit{Proc. 2017 IEEE International Symposium on Information Theory (ISIT)}, Aachen, Germany, June. 2017, pp. 988--992.
%%=======================================


%%======
\bibitem{ref_Shilling-attack-detection}
J. Cao, Z. Wu, B. Mao, and Y. Zhang.
''Shilling attack detection utilizing semi-supervised learning method for collaborative recommender system,''
\textit{World Wide Web Journal: Internet and Web Information Systems},
Vol. 16, no. 5--6, pp. 729--748, 2013.

\bibitem{ref_A-recommendation-engine}
G. Zhu, J. Cao, C. Li, Z. Wu.
``A recommendation engine for travel products based on topic sequential patterns,'' \textit{Multimedia Tools and Application},
vol. 76, no. 16, pp. 17595--17612, 2017.

\bibitem{ref_Hybrid-collaborative}
J. Cao, Z. Wu, Y. Wang and Y. Zhuang,
``Hybrid collaborative filtering algorithm for bidirectional web service recommendation,'' \textit{Knowledge and Information Systems},
vol. 36, no. 3, pp. 607--627, 2013.


%%%%%

%\bibitem{ref_Counterterrorism-systems}
%A. Zieba, ''Counterterrorism systems of spain and poland: Comparative studies,''
%\textit{Przeglad Politologiczny}, vol. 3, pp. 65--78, 2015.

%\bibitem{A-mathematical-theory}
%C. E. Shannon, ``A mathematical theory of Communicaiton,''
%\textit{The Bell Syst. Tech. J.}, Vol.27, pp.379--423, 623-656, July--Oct. 1948.
%
%\bibitem{On-measures-of-entropy}
%A. Renyi, ``On measures of entropy and information,''
%in \textit{Proc. 4th Berkeley Symp. Math. Statist. and Probability}, vol. 1. 1961, pp. 547--561.
%
%\bibitem{Elements-of-information-theory-2nd-edition}%8
%T. M. Cover and J. A. Thomas. \textit{Elements of information theory 2nd edition}, Wiley Series in Telecommunications and Signal Processing, Wiley InterScience, 2006.
%
%\bibitem{Universal-outlier-hypothesis-testin}
%Y. Li, S. Nitinawarat, and V. V. Veeravalli,
%``Universal outlier hypothesis testin,''
%\textit{IEEE trans. inf. theory}, vol. 60, no. 7, pp. 4066--4082, July 2014.

%%%%
\bibitem{ref_Message-Importance-Measure-and-Its-Application-to-Minority-Subset-Detection-in-Big-Data}
P. Fan, Y. Dong, J. X. Lu, and S. Y. Liu, ``Message importance measure and its application to minority subset detection in big data,'' in \textit{Proc. IEEE Globecom Workshops (GC Wkshps)}, Washington D.C., USA, Dec. 2016, pp. 1--6.

\bibitem{ref_Recognizing-Information-Feature-Variation}
R. She, S. Liu and P. Fan. ``Recognizing information feature variation: message importance transfer measure and its applications in big Data,''
\textit{Entropy}, vol. 20, no. 6, pp 1--22, May 2018.

\bibitem{Matching-users-preference}
{S. Liu, Y. Dong, P. Fan, R. She and S. Wan}.
{``Matching users' preference under target revenue constraints in data recommendation systems,''}
\textit{Entropy}, vol. {21}, no. {2}, pp. {205}, Feb. {2019}.

\bibitem{ref_Differential-message-importance-measure}
S. Liu, R. She, and P. Fan,
``Differential message importance measure: a new approach to the required sampling number in big data structure characterization,''
\textit{IEEE Access}, vol. 6, pp. 42851--42867, July 2018.

\bibitem{ref_Minor-probability-events}
S. Wan , J. Lu , P. Fan, and K. B. Letaief,
``Minor probability events' detection in big data: an integrated approach with bayes detection and MIM,''
\textit{ IEEE Commun. Lett.}, vol. 23, no. 3, pp. 418--421, Mar. 2019.


\bibitem{ref_Focusing-on-a-Probability-Element}%\label{ref16}
R. She, S. Y. Liu, Y. Q. Dong, and P. Fan, ``Focusing on a probability element: parameter selection of message importance measure in big data,'' in \textit{Proc. IEEE International Conference on Communications (ICC)}, Paris, France, May. 2017, pp. 1--6.

%\bibitem{Fraudulent-Financial-Reporting}
%M. S. Beasley, J. V. Carcello, D. R. Hermanson and P. D. Lapides,
%``Fraudulent financial reporting: Consideration of industry traits and corporate governance mechanisms,''
%\textit{Accounting Horizons}, vol. 14, no. 4, pp. 441--454, Dec. 2000.
%
%\bibitem{Counterterrorism-systems}
%A. Zieba, ``Counterterrorism systems of spain and poland: Comparative studies,''
%\textit{Przeglad Politologiczny}, vol. 3, pp. 65--78, 2015.

\bibitem{ref_Non-parametric-Message-Important-Measure}%\label{ref}
S. Liu, R. She, P. Fan, and J. Lu, ``Non-parametric message important measure: Compressed storage design for big data in wireless communication systems,'' in \textit{Proc. IEEE Asia-Pacific Conference on Communications (APCC)}, Perth, Australia, Dec. 2017, pp. 1--6.

\bibitem{ref_Storage-Code-Design-and-Transmission}
S. Liu, R. She, P. Fan, K. B. Letaief, ``Non-parametric Message Importance Measure: Storage Code Design and Transmission Planning for Big Data,''
\textit{IEEE Trans. Commun.}, pp. 1--1, Jun. 2018.


%%========================================
\bibitem{ref_Big-data-related-technologies}
M. Chen, S. Mao, Y. Zhang, and V. C. Leung, Big data: related technologies, challenges and future prospects. Springer, 2014.

\bibitem{ref_Spatiotemporal-stochastic-modeling}
M. Gharbieh, H. ElSawy, A. Bader, and M. Alouini, ``Spatiotemporal stochastic modeling of IoT enabled cellular networks: scalability and stability analysis,'' \textit{IEEE Trans. Commun.}, vol. 65, no. 8, pp. 3585--3600, Aug. 2017.

\bibitem{ref_Wireless-communications-in-the-era}
S. Bi, R. Zhang, Z. Ding, and S. Cui, ``Wireless communications in the era of big data,'' \textit{IEEE Commun. Mag.}, vol. 53, no. 10, pp. 190--199, Oct. 2015.

%%===========================

%%========================
\bibitem{ref_Lossy-compression-for-compute-and-forward}
I. E. Aguerri and A. Zaidi, ``Lossy compression for compute-and-forward in limited backhaul uplink multicell processing,'' \textit{IEEE Trans. Commun.}, vol. 64, no. 12, pp. 5227--5238, Dec. 2016.

\bibitem{ref_Distributed-distortion-optimization-for}
T. Cui, L. Chen, and T. Ho, ``Distributed distortion optimization for correlated sources with network coding,'' \textit{IEEE Trans. Commun.}, vol. 60, no. 5, pp. 1336--1344, May 2012.

\bibitem{ref_Block-and-sliding-block-lossy}
S. Jalali and T. Weissman, ``Block and sliding-block lossy compression via mcmc,'' \textit{IEEE Trans. Commun.}, vol. 60, no. 8, pp. 2187--2198, Aug. 2012.

\bibitem{ref_On-the-rate-distortion-function}
A. Sechelea, A. Munteanu, S. Cheng, and N. Deligiannis, ``On the rate-distortion function for binary source coding with side information,'' \textit{IEEE Trans. Commun.}, vol. 64, no. 12, pp. 5203--5216, Dec. 2016.

%%======================================
\bibitem{ref_Fifty-years-of}
S. Verdu, ``Fifty years of shannon theory,'' \textit{IEEE Trans. Inf. Theory}, vol. 44, no. 6, pp. 2057--2078, 1998.

\bibitem{ref_Elements-of-information-theory-2nd-edition}%8
T. M. Cover and J. A. Thomas. \textit{Elements of information theory 2nd edition}, Wiley Series in Telecommunications and Signal Processing, Wiley InterScience, 2006.

\bibitem{ref_Renyi-divergence-and-kullback-leibler}
T. V. Erven and P. Harremoes, ``Renyi divergence and kullback-leibler divergence,'' \textit{IEEE Trans. Inf. Theory}, vol. 60, no. 7, pp. 3797--3820, Jun. 2014.

\bibitem{ref_Information-theory-and-an-extension}
H. Akaike, ''Information theory and an extension of the maximum likelihood principle,'' in \textit{Selected Papers of Hirotugu Akaike. Springer}, 1998, pp. 199--213.

%%%======================================================
\bibitem{ref_On-linear-unequal-error}
B. Masnick and J. Wolf, ``On linear unequal error protection codes,''
\textit{IEEE Trans. Inf. Theory},
vol. 3, no. 4, pp.600--607, Oct. 1967.

\bibitem{ref_Expanding-window-fountain}
D. Sejdinovic, D. Vukobratovic, A. Doufexi, V. Senk and R.J. Piechocki,
``Expanding window fountain codes for unequal error protection,''
\textit{IEEE Trans. Commun.}, vol. 57, no. 9, pp. 2510--3526, Sep. 2009.

\bibitem{ref_Unequal-error-protection}
K. Sun and D. Wu,
``Unequal error protection for video streaming using delay-aware fountain codes,''
in \textit{Proc. IEEE International Conference on Communications (ICC)},
Paris, France, May. 2017, pp. 1--6.

%%%=======================================================
%\bibitem{ref_A-measure-of-asymptotic-efficiency}
%H. Chernoff, ``A measure of asymptotic efficiency for tests of a hypothesis based on a sum of observations,'' \textit{Ann. Mathe. Stat.}, vol. 23, pp. 493--507, Dec. 1952.
%
%\bibitem{ref_Differential-Geometrical-Methods}
%S. Amari, \textit{Differential-geometrical methods in statistics}, Springer Science \& Business Media, Tokyo, 2012.
%
%\bibitem{ref_Robust-paramater-estimation}
%H. Fujisawa and S. Eguchi, ``Robust paramater estimation with a small bias against heavy contamination,'' \textit{ J. Multivariate. Anal.}, vol. 99, no. 9, pp. 2053--2081, Oct. 2008.

\bibitem{ref_Universal-outlier-hypothesis-testin}
Y. Li, S. Nitinawarat, and V. V. Veeravalli,
``Universal outlier hypothesis testin,''
\textit{IEEE trans. inf. theory}, vol. 60, no. 7, pp. 4066--4082, July 2014.

%%============================================
\bibitem{ref_Cost-Sensitive-encoding}
K. Lo and H. Lin,
``Cost-Sensitive encoding for label space dimension reduction algorithms on multi-label classification''
in \textit{Proc. Conference on Technologies and Applications of Artificial Intelligence (TAAI)},
Taipei, Taiwan, Dec. 2017, pp. 136--141.

\bibitem{ref_Discriminant-analysis-based-dimension}
W. Li, F. Feng, H. Li and Q. Du
''Discriminant analysis-based dimension reduction for hyperspectral image classification: a survey of the most recent advances and an experimental comparison of different techniques,''
\textit{IEEE Geosci. Remote Sens. Mag.},
vol. 6, no. 1, pp. 15--34, Mar. 2018.

\bibitem{ref_Local-deep-feature}
J. Zhang, J. Yu, and D. Tao,
``Local deep-feature alignment for unsupervised dimension reduction,''
\textit{IEEE Trans. Image Process.},
vol. 27, no. 5, pp. 2420--2432, Feb. 2018.

%%%=========================================
\bibitem{ref_A-Survey-of-Data-Mining}
A. L. Buczak and E. Guven, ``A Survey of Data Mining and Machine Learning Methods for Cyber Security Intrusion Detection,''
\textit{IEEE communications surveys \& tutorials}, vol. 18, no. 2, May 2016.

\bibitem{ref_A-Deep-Learning-Approach}
C. Yin , Y. Zhu, J. Fei, and X. He, ``A Deep Learning Approach for Intrusion Detection Using Recurrent Neural Networks,''
\textit{ IEEE Internet Things J.}, vol. 5, pp. 21954--21961, Nov. 2017.

\bibitem{ref_Toward-a-Gaussian}
X. Yang, P. Zhao, X. Zhang, J. Lin, and W. Yu,
``Toward a Gaussian mixture model-based detection scheme against data integrity attacks in the smart grid,''
\textit{ IEEE Internet Things J.}, vol. 4, no. 1, pp. 147--161, Feb. 2017.

%%===========================================================================

%%==========================================================
\bibitem{ref_A-characterization-of-limiting}
J. Hajek, ``A characterization of limiting distributions of regular estimates,''
\textit{Zeitschrift Wahrscheinlichkeitstheorie Verwandte Gebiete},
vol. 14, no. 4, pp. 323--330, 1970.

\bibitem{ref_Local-asymptotic-minimax}
J. Hajek,
``Local asymptotic minimax and admissibility in estimation,''
in \textit{Proc. 6th Berkeley Symp. Math. Statist. Probab.},
vol. 1, 1972, pp. 175--194.

\bibitem{ref_Asymptotic-Methods-in-Statistical}
L. Le Cam,
\textit{Asymptotic Methods in Statistical Decision Theory},
New York, NY, USA: Springer, 1986.

\bibitem{ref_Estimating-entropy-on-m-bins}
L. Paninski,
``Estimating entropy on m bins given fewer than m samples,''
\textit{ IEEE Trans. Inf. Theory}, vol. 50, no. 9, pp. 2200--2203, 2004.

%%=======a) esitmation======================================
\bibitem{ref_Minimax-estimation-of-discrete}
Y. Han, J. Jiao, and T. Weissman,  ``Minimax estimation of discrete distributions,''
in \textit{Proc. 2015 IEEE International Symposium on Information Theory (ISIT)}, Hong Kong, June. 2015, pp. 2291--2295.

\bibitem{ref_Minimax-estimation-of-discrete-distributions-under}
Y. Han, J. Jiao, and T. Weissman,
``Minimax estimation of discrete distributions under $\ell_{1}$ loss,'' \textit{ IEEE Trans. Inf. Theory}, vol. 61, no. 11, pp. 6343--6354, Sep. 2015.

\bibitem{ref_Minimax-risk-over}
D. L. Donoho and I. M. Johnstone, ``Minimax risk over $\ell_p$-balls for
$\ell_p$-error,'' \textit{Probab. Theory Rel. Fields}, vol. 99, no. 2, pp. 277--303, 1994.

\bibitem{ref_Ideal-spatial-adaptation}
D. L. Donoho and J. M. Johnstone, ``Ideal spatial adaptation
by wavelet shrinkage,'' \textit{Biometrika}, vol. 81, no. 3, pp. 425--455,
1994.

\bibitem{ref_Beyond-histograms}
Diakonikolas, ``Beyond histograms: Structure and distribution estimation,'' in \textit{STOC'14 Workshop}, New York, USA, May. 2014.

%%=======b) estimation====================

\bibitem{ref_Minimax-estimation-of-information-measures}
J. Jiao, K. Venkat, Y. Han, and T. Weissman, ``Minimax estimation of information measures,'' in
\textit{Proc. 2015 IEEE International Symposium on Information Theory (ISIT)},
Hong Kong, June. 2015, pp. 2296--2300.

\bibitem{ref_Minimax-estimation-of-functionals-of-discrete-distributions}
J. Jiao, K. Venkat, Y. Han, and T. Weissman,
``Minimax estimation of functionals of discrete distributions''
\textit{ IEEE Trans. Inf. Theory}, vol. 61, no. 5, pp. 2835--2885 , May. 2015.

\bibitem{ref_Mutualinformation-based-registration-of-medical-images}
J. P. W. Pluim, J. B. A. Maintz, and M. A. Viergever, ``Mutual information-based registration of medical images: A survey,'' \textit{IEEE Trans. Med. Imag.}, vol. 22, no. 8, pp. 986--1004, Aug. 2003.

\bibitem{ref_Alignment-by-maximization-of-mutual-information}
P. Viola and W. M. Wells, III, ``Alignment by maximization of mutual information,'' \textit{Int. J. Comput. Vis.}, vol. 24, no. 2, pp. 137--154, Sep. 1997.

\bibitem{ref_Mutual-information-analysis}
L. Batina, B. Gierlichs, E. Prouff, M. Rivain, F. X. Standaert, and N. V. Charvillon, ``Mutual information analysis: A comprehensive study,''
\textit{J. Cryptol.}, vol. 24, no. 2, pp. 269--291, Apr. 2011.

\bibitem{ref_Maximum-Likelihood-Estimation-of-information-measures}
J. Jiao, K. Venkat, Y. Han, and T. Weissman,
``Maximum likelihood estimation of functionals of discrete distributions,''
\textit{IEEE Trans. Inf. Theory}, vol. 63, no. 10, pp. 6774--6798, Oct. 2017.

%%====c) estimation=================================
\bibitem{ref_Adaptive-estimation-of-Shannon-entropy}
Y. Han, J. Jiao, and T. Weissman,
``Adaptive estimation of Shannon entropy,'' \textit{Proc. 2015 IEEE International Symposium on Information Theory (ISIT)},
Hong Kong, June. 2015, pp. 1372--1376.

\bibitem{ref_Optimal-entropy-estimation-on-large-alphabets-via-best-polynomial-approximation}
Y. Wu and P. Yang,
``Optimal entropy estimation on large alphabets via best polynomial approximation,''
\textit{Proc. 2015 IEEE International Symposium on Information Theory (ISIT)},
Hong Kong, June. 2015, pp. 824--828.

\bibitem{ref_Does-dirichlet-prior-smoothing}
Y. Han, J. Jiao, and T. Weissman,
``Does Dirichlet prior smoothing solve the Shannon entropy estimation problem?''
\textit{Proc. 2015 IEEE International Symposium on Information Theory (ISIT)},
Hong Kong, June. 2015, pp. 1367--1371.

\bibitem{ref_Ensemble-estimators-for-multivariate-entropy-estimation}
K. Sricharan, D. Wei, and A. O. Hero,
``Ensemble estimators for multivariate entropy estimation,''
\textit{ IEEE Trans. Inf. Theory}, vol. 59, no. 7, pp. 4374--4388, Mar. 2013.

%%==========d) estimation==========================================
\bibitem{ref_Estimation-of-KL-divergence-between-large-alphabet-distributions}
Y. Bu, S. Zou, Y. Liang, and V. V. Veeravalli,
``Estimation of KL divergence between large-alphabet distributions,''
\textit{Proc. 2016 IEEE International Symposium on Information Theory (ISIT)},
Barcelona, Spain, July. 2016, pp. 1118--1122.

\bibitem{ref_Estimating-the-unseen}
G. Valiant and P. Valiant, ``Estimating the unseen: an n/log (n)-sample estimator for entropy and support size, shown optimal via new clts,`` in \textit{Proc. of the 43rd annual ACM symposium on Theory of computing}. San Jose, USA, June, 2011, pp. 685--694.

\bibitem{ref_Minimax-rates-of-entropy-estimation}
Y. Wu and P. Yang, ``Minimax rates of entropy estimation on large alphabets via best polynomial approximation,''
\textit{ IEEE Trans. Inf. Theory}, vol. 62, no. 6, pp. 3702--3720, Mar. 2016.

\bibitem{ref_Divergence-estimation-of-continuous-distributions}
Q. Wang, S. R Kulkarni, and S. Verdu, ``Divergence estimation of continuous distributions based on data-dependent partitions,''
\textit{ IEEE Trans. Inf. Theory}, vol. 51, no. 9, pp. 3064--3074, Aug. 2005.

\bibitem{ref_Divergence-estimation-for-multidimensional-densities}
Q. Wang, S. R Kulkarni, and S. Verdu,``Divergence estimation for multidimensional densities via k-nearest-neighbor distances,'' \textit{ IEEE Trans. Inf. Theory}, vol. 55, no. 5, pp. 2392--2405, May. 2009.

\bibitem{ref_Estimating-divergence-functionals-and-the-likelihood-ratio}
X. Nguyen, M. J Wainwright, and M. Jordan, ''Estimating divergence functionals and the likelihood ratio by convex risk minimization,'' \textit{ IEEE Trans. Inf. Theory},
vol. 56, no. 11, pp. 5847--5861, Oct. 2010.

\bibitem{ref_Minimax-rate-optimal-estimation-of-KL-divergence}
Y. Han, J. Jiao, and T. Weissman,
''Minimax rate-optimal estimation of KL divergence between discrete distributions,''
in \textit{Proc. 2016 International Symposium on Information Theory and Its Applications (ISITA)}, Monterey, USA, Oct. 2016, pp. 256--260.

\bibitem{ref_Ensemble-estimation-of-multivariate-f-divergence}
K. R. Moon and A. O. Hero,
''Ensemble estimation of multivariate f-divergence,'' \textit{Proc. 2014 IEEE International Symposium on Information Theory(ISIT)}, Honolulu, USA, June. 2014, pp. 356--360.

%%======================================================================
\bibitem{ref_Principal-component-analysis}
I. T. Jolliffe, \textit{Principal Component Analysis}. New York, NY, USA: Springer-Verlag, 1986.

\bibitem{ref_Independent-component-analysis}
A. Hyvarinen, J. Karhunen, and E. Oja, \textit{Independent Component Analysis}. New York, NY, USA: Wiley, 2001.

\bibitem{ref_Statistical-Models-Theory-and-Practice}
D. A. Freedman, \textit{Statistical Models: Theory and Practice}. Cambridge, U.K.: Cambridge Univ. Press, 2005.

\bibitem{ref_K-means-clustering-via-principal-component-analysis}
C. Ding and X. He, ``K-means clustering via principal component analysis,'' in \textit{Proc. 21st ICML}, New York, NY, USA, 2004, pp. 29--36.

\bibitem{ref_A-new-approach-to-linear-filtering}
R. E. Kalman, ''A new approach to linear filtering and prediction problems,'' \textit{J. Fluids Eng.}, vol. 82, no. 1, pp. 35--45, Mar. 1960.

\bibitem{ref_Probabilistic-Graphical-Models}
D. Koller and N. Friedman, \textit{Probabilistic Graphical Models: Principles and Techniques}. Cambridge, MA, USA: MIT Press, 2009.

\bibitem{ref_Linear-information-coupling-problems}
S. L. Huang and L. Zheng, ''Linear information coupling problems,'' in \textit{Proc. 2012 IEEE International Symposium on Information Theory (ISIT)}, Cambridge, USA, Jul. 2012, pp. 1029--1033.

\bibitem{ref_The-Linear-information}
S. L. Huang and Li. Zheng. (2014).
The Linear information coupling problems.
[Online].
https://arxiv.org/abs/1406.2834


%%=======================================================================
\bibitem{ref_Universal-estimation-of-directed-information}
J. Jiao, H. H. Permuter, L. Zhao, Y. Kim, and T. Weissman,
''Universal estimation of directed information''
\textit{ IEEE Trans. Inf. Theory}, vol. 59, no. 10, pp. 6220--6242 , July. 2013.

\bibitem{ref_Investigating-causal-relations}
C.Granger, ''Investigating causal relations by econometric models and cross-spectral methods,'' \textit{Econometrica}, vol. 37, no. 3, pp. 424--438, 1969.

\bibitem{ref_Using-directed-information-to-build-biologically}
A. Rao, A. O. Hero, D. J. States, and J. D. Engel, ''Using directed information to build biologically relevant influence networks,'' \textit{J. Bioinf. Comput. Biol.}, vol. 6, no. 3, pp. 493--519, 2008.

\bibitem{ref_Universal-divergence-estimation}
H. Cai, S. R. Kulkarni, and S. Verdu, ''Universal divergence estimation for finite-alphabet sources,'' \textit{IEEE Trans. Inf. Theory}, vol. 52, no. 8, pp. 3456--3475, Aug. 2006.

\bibitem{ref_Thecontext-tree-weight-ingmethod}
F. M. J. Willems, Y. M. Shtarkov, and T. J. Tjalkens, ''The context-tree weight ingmethod: Basic properties,'' \textit{IEEE Trans. Inf. Theory}, vol. 41, no. 3, pp. 653--664, May 1995.

\bibitem{ref_Universal-directed-information}
L. Zhao, Y. H. Kim, H. H. Permuter, and T. Weissman, ''Universal estimation of directed information,'' in \textit{Proc. IEEE Int. Symp. Inf. Theory}, Austin, USA, June. 2010, pp. 230--234.

\bibitem{ref_Causality-feedback-and-directed}
J.Massey, ''Causality, feedback and directed information,'' in \textit{Proc. Int. Symp. Inf. Theory Applic. (ISITA-90)}, Nov. 1990, pp. 303--305.

\bibitem{ref_The-bidirectional-communication-theory}
H. Marko, ''The bidirectional communication theory: A generalization ofinformationtheory,'' \textit{IEEE Trans. Commun.}, vol. 21, pp. 1335--1351, 1973.

\bibitem{ref_Interpretations-of-directed-information}
H. H. Permuter, Y. Kim, and T. Weissman,
''Interpretations of directed information in portfolio theory, data compression, and hypothesis testing,'' \textit{ IEEE Trans. Inf. Theory}, vol. 57, no. 6, pp. 3248--3259 , May. 2011.

\bibitem{ref_Directed-Information-for-Channels-With-Feedback}
G. Kramer, ''Directed Information for Channels With Feedback,'' Ph.D. dissertation, Swiss Fed. Inst. Technol. (ETH), Zurich, Switzerland, 1998.

\bibitem{ref_A-coding-theorem}
Y. H. Kim, ''A coding theorem for a class of stationary channels with feedback,'' \textit{IEEE Trans. Inf. Theory.}, vol. 25, no. 4, pp. 1488--1499, Apr. 2008.

\bibitem{ref_Estimating-the-directed-information}
C. J. Quinn, T. P. Coleman, N. Kiyavash, and N. G. Hatsopoulos, ``Estimating the directed information to infer causal relationships in ensemble neural spike train recordings,'' \textit{J. Comput. Neurosci.}, pp. 1-28, 2010.

\bibitem{ref_Mapping-information-flow}
M. Lungarella and O. Sporns, ``Mapping information flow in sensorimotor networks,'' \textit{PLoS Comput. Biol.}, vol. 2, no. 10, pp. 1301--1312, Oct. 2006.

\bibitem{ref_MAchieving-the-Gaussian-rate}
R. Zamir, Y. Kochman, and U. Erez, ``Achieving the Gaussian rate distortion function by prediction,'' \textit{IEEE Trans. Inf. Theory}, vol. 54, no. 7, pp. 3354--3364, Jul. 2008.

%%==============Section prob derivation======================================
%\vspace{-3mm}
%\begin{shaded}
%\vspace{-3mm}
\bibitem{ref_A-unified-framework-for-event}
J. Kwon and K. M. Lee,
``A unified framework for event summarization and rare event detection,''
in \textit{Proc. 2012 IEEE Conference on Computer Vision and Pattern Recognition (IEEE CVPR)},
Providence, USA, Jun. 2012, pp. 1266--1273.

\bibitem{ref_Semi-Markov-switching}
I. Melnyk, A. Banerjee, B. Matthews, and N. Oza,
``Semi-Markov switching vector autoregressive model-based anomaly detection in aviation systems,''
in \textit{Proc. ACM SIGKDD International Conference on Knowledge Discovery and Data Mining (KDD)}, San Francisco, USA, Aug, 2016, pp. 1065--1074.

\bibitem{ref_Spatio-temporal-network-anomaly}
I. Paschalidis and G. Smaragdakis,
``Spatio-temporal network anomaly detection by assessing deviations of empirical measures,'' \textit{IEEE/ACM Trans. Netw.}, vol. 17, no.3, pp. 685--697, Jun. 2009.

\bibitem{ref_Group-anomaly-detection}
L. Xiong, B. Poczos, and J. Schneider.
``Group anomaly detection using flexible genre models,''
in \textit{Proc. Advances in neural information processing systems (NIPS)},
Granada, Spain, Dec. 2011, pp. 1071--1079.

\bibitem{ref_Detecting-and-classifying}
W. C. Young, J. E. Blumenstock, E. B. Fox, and T. H. McCormick,
``Detecting and classifying anomalous behavior in spatiotemporal network data,''
in \textit{Proc. KDD workshop on learning about emergencies from social information (KDD--LESI)}, New York, USA, Jun. 2014, pp, 29--33.

%%%===================================

\bibitem{ref_On-mining-anomalous}
L. X. Pang, S. Chawla, W. Liu, and Y. Zheng, .
``On mining anomalous patterns in road traffic streams,''
in \textit{Proc. International Conference on Advanced Data Mining and Applications (ADMA)}.
Nanjing, China, Dec. 2011, pp. 237--251.

\bibitem{ref_Detecting-collective-anomalies}
Y. Zheng, H. Zhang, and Y. Yu,
``Detecting collective anomalies from multiple spatio-temporal datasets across different domains,''
in \textit{Proc. ACM SIGSPATIAL International Conference on Advances in Geographic Information Systems},
Bellevue, USA , Nov. 2015, pp. 1--10.

\bibitem{ref_Spatio-Temporal-Event-Detection-Using}
J. Yin, D. H. Hu, and Q. Yang,
``Spatio-temporal event detection using dynamic conditional random fields,''
in \textit{Proc. International Jont Conference on Artifical Intelligence (IJCAI)},
Pasadena, USA, Jul. 2009, pp. 1321--1327.

\bibitem{ref_Anomalous-event-detection-on-large-scale}
A. Witayangkurn, T. Horanont, Y. Sekimoto, and R. Shibasaki.
``Anomalous event detection on large-scale gps data from mobile phones using hidden markov model and cloud platform,''
in \textit{Proc. ACM conference on Pervasive and ubiquitous computing adjunct publication (UbiComp)},
Zurich, Switzerland, Sep. 2013, pp. 1219--1228.

%%%=================================

\bibitem{ref_Multicriteria-Similarity-Based}
K. Hsiao, K. S. Xu, J. Calder, and A. O. Hero,
``Multicriteria similarity-based anomaly detection using pareto depth analysis,''
\textit{IEEE Trans. Neural Netw. Learn. Syst.},
vol. 27, no. 6, pp. 1307--1321, Jun. 2016.

\bibitem{ref_Visual-Analysis-of-Time-Series}
M. Steiger, J. Bernard, S. Mittelstadt, H. L. Tieke, D. Keim, T. May, and J. Kohlhammer,
``Visual analysis of time-series similarities for anomaly detection in sensor networks,''
\textit{Computer Graphics Forum}, vol. 33, no. 3, pp. 401--410, Jun. 2014.

\bibitem{ref_Temporal-outlier-detection}
X. Li, Z. Li, J. Han, and J. Lee,
``Temporal outlier detection in vehicle traffic data,''
in \textit{Proc. IEEE International Conference on Data Engineering (ICDE)},
Shanghai, China, 29 Mar. 2009, pp. 1319--1322.

\bibitem{ref_Detecting-rare-events-using-Kullback¨CLeibler}
J. Xu, S. Denman, C. Fookes, and S. Sridharan,
``Detecting rare events using Kullback¨CLeibler divergence: A weakly supervised approach,''
\textit{Expert Systems With Applications},
vol. 54, pp. 13--28, Jul. 2016.

\bibitem{ref_Anomaly-Detection-and-Localization}
W. Li, V. Mahadevan, and N. Vasconcelos,
``Anomaly Detection and Localization in Crowded Scenes,''
\textit{IEEE Trans. Pattern Anal. Mach. Intell.}
vol. 36, no. 1, pp. 18--32, Jan. 2014.
%\end{shaded}



%%%============Lizhong=============================
%\bibitem{Exact-Bayesian-structure-discovery}
%M. Koivisto and K.Sood, ``Exact Bayesian structure discovery in Bayesian networks,'' \textit{J. Mach. Learn. Res.}, vol. 5, pp. 549--573, May. 2004.
%
%\bibitem{Learning-causal-relations}
%Z. Wang and L. Chan, ``Learning causal relations in multivariate time series data,'' \textit{ACM Trans. Intell. Syst. Technol.}, vol. 3, no. 4, p. 76, Sep. 2012.
%
%\bibitem{Big-data-analysis-with}
%A. Sandryhalia and J. M. F. Moura, ``Big data analysis with signal processing on graphs,'' \textit{IEEE Signal Process. Mag.}, vol. 31, no. 5, pp. 80--90, Sep. 2014.
%
%\bibitem{Multiuser-Detection}
%S. Verdú, \textit{Multiuser Detection}. Cambridge, U.K.: Cambridge Univ. Press, 1998.
%
%\bibitem{Broadband-MIMO-OFDM-wireless}
%G. Stuber et al., ``Broadband MIMO-OFDM wireless communications,'' \textit{IEEE J. Sel. Areas Commun.}, vol. 92, no. 2, pp. 271--294, Feb. 2004.
%
%%\bibitem{Graphical-models-exponential-families}
%%M. J. Wainwright and M. I. Jordan, ``Graphical models, exponential families, variational inference,'' \textit{Found. Trends Mach. Learn.}, vol. 1, no. 1/2, pp. 1--305, Jan. 2008.
%
%\bibitem{From-technological-networks-to}
%K. C. Chen, M. Chiang, and H. V. Poor, ``From technological networks to social networks,'' \textit{IEEE J. Sel. Areas Commun.}, vol. 31, no. 9, pp. 548--572, Sep. 2013.
%
%\bibitem{Communication-theoretic-data-analytics}
%K. Chen, S. Huang, L. Zheng, and H. V. Poor,
%``Communication theoretic data analytics,'' \textit{IEEE Journal on Selected Areas in Communications}, vol. 33, no. 4, pp. 663--675, Apri. 2015,
%
%\bibitem{Communication-theoretic-inference-on}
%K. Chen, B. Mankir, S. Huang, L. Zheng, and H. V. Poor,
%``Communication theoretic inference on heterogeneous data,'' in \textit{Proc.
%2016 IEEE International Conference on Communications (ICC)}, Kuala Lumpur, Malaysiapp, May. 2016,
%pp. 1--6.
%
%\bibitem{Distributed-detection-with-multiple}
%R. S. Blum, S. A. Kassam, and H. V. Poor, ``Distributed detection with multiple sensors: Part II,'' \textit{Proc. IEEE}, vol. 85, no. 1, pp. 64--79, Jan. 1997.

%%==========smart city=============
%\bibitem{World's-population-increasingly}
%United Nations, ``World's population increasingly urban with more than half living in urban areas,'' 2014. [Online]. Available:
%http://www.un.org/en/development/desa/news/population/
%worldurbanization-prospects-2014.html

\bibitem{ref_A-Survey-on-data}
A. Gharaibeh, M. A. Salahuddin, S. J. Hussini, A. Khreishah, I. Khalil, M. Guizani, and A. A. Fuqaha,
``Smart cities: A Survey on data management, security, and enabling technologies,''
\textit{IEEE Communications Surveys \& Tutorials}, vol. 19, no. 4, pp. 2456--2501, Aug. 2017.

\bibitem{ref_An-architecture-for-the}
V. G. Font, C. Garrigues, and H. R. Pous,
``An architecture for the analysis and detection of anomalies in smart city WSNs,''
in \textit{Proc. 2015 IEEE First International Smart Cities Conference (ISC2)}, Guadalajara, Mexico, Oct. 2015, pp. 1--6.

\bibitem{ref_An-anomaly-detection-in}
R. Jain, and H. Shah,
``An anomaly detection in smart cities modeled as wireless sensor network,'' in  \textit{
Proc. 2016 International Conference on Signal and Information Processing (IConSIP)}, Nanded, India, Oct. 2016, pp. 1--5.

\bibitem{ref_IoT-driven-automated-object}
L. Hu and Q. Ni
``IoT-driven automated object detection algorithm for urban surveillance systems in smart cities, '' \textit{IEEE Internet of Things Journal}, vol. pp, no. 99, pp. 1--8, May. 2017.
%%======Spatial-Temporal Data============
%\vspace{-1mm}
%\begin{shaded}
%\vspace{-3mm}
\bibitem{ref_Detecting-Urban-Anomalies-Using}
H. Zhang, Y. Zheng, and Y. Yu.
``Detecting urban anomalies using multiple spatio-temporal data sources,''
\textit{Proceedings of the ACM on Interactive, Mobile,Wearable and Ubiquitous Technologies},
vol. 2, no. 1, pp. 54:1--18, Mar. 2018.

\bibitem{ref_Spatio-temporal-Anomaly}
Q. Wang, W. Lv, and B. Du,
``Spatio-temporal anomaly detection in traffic data,''
in \textit{Proc. 2nd International Symposium on Computer Science and Intelligent Control (ISCSIC '18)}, Stockholm, Sweden, Sep. 2018, pp. 1--5.

\bibitem{ref_Uapd}
X. Wu, Y. Dong, C. Huang, J. Xu, D. Wang, and N. V. Chawla,
``Uapd: Predicting urban anomalies from spatial-temporal data,''
in \textit{Joint European Conference on Machine Learning and Knowledge Discovery in Databases}, Springer, Cham., Sep. 2017, pp. 622--638.
%\end{shaded}

%%===black hole===========================

\bibitem{ref_Detecting-urban-black-holes}
L. Hong, Y. Zheng, D. Yung, J. Shang, and L. Zou,
``Detecting urban black holes based on human mobility data,'' in \textit{Proc. SIGSPATIAL'15}, Bellevue, USA, Nov. 2015.

\bibitem{ref_A-context-aware-collaborative}
L. Jin, Z. Feng, and L. Feng,
``A context-aware collaborative filtering approach for urban black holes detection,''
in \textit{Proc. CIKM'16}, Indianapolis, USA, Oct. 2016, pp. 2137--2142.

\bibitem{ref_Scalable-graph-clustering-using}
V. Satuluri and S. Parthasarathy, ``Scalable graph clustering using stochastic flows: applications to community discovery,'' in \textit{Proc. of KDD}, ACM, pp. 737--746, 2009.

\bibitem{ref_Graph-clustering-based-on}
Y. Zhou, H. Cheng, and J. X. Yu, ``Graph clustering based on structural/attribute similarities,'' in \textit{Proc. of VLDB Conference}, 2009.

\bibitem{ref_On-modularity-clustering}
U. Brandes, D. Delling, M. Gaertler, R. Gorke, M. Hoefer, Z. Nikoloski, and D. Wagner,
``On modularity clustering,'' \textit{IEEE Trans. on Knowledge and Data Engineering}, vol. 20, no. 2, pp. 172--188, 2008.

\bibitem{ref_Spotting-significant-changing-subgraphs}
Z. Liu, J. X. Yu, Y. Ke, and X. Lin, ``Spotting significant changing subgraphs in evolving graphs,'' in \textit{Proc. of ICDM}, 2008.

\bibitem{ref_Constraint-based-pattern-mining}
C. Robardet, ``Constraint-based pattern mining in dynamic graphs,'' in \textit{ Proc. of ICDM}, 2009.

\bibitem{ref_Effective-density-queries-on}
C. S. Jensen, D. Lin, B. C. Ooi, and R. Zhang, ``Effective density queries on continuously moving objects,'' in \textit{Proc. of ICDE}, 2006.

\bibitem{ref_Mining-relaxed-temporal-moving}
Z. Li, B. Ding, J. Han, and R. Kays, ``Swarm: Mining relaxed temporal moving object clusters,'' \textit{PVLDB}, vol. 3, no. 1-2, pp. 723--734, 2010.

\bibitem{ref_On-discovery-of-gathering}
K. Zheng, Y. Zheng, N. J. Yuan, and S. Shang, ``On discovery of gathering patterns from trajectories,'' in \textit{Proc. of ICDE}, 2013.

\bibitem{ref_Identifying-attributing-and-describing}
M. Mathioudakis, N. Bansal, and N. Koudas, ``Identifying, attributing and describing spatial bursts,'' \textit{PVLDB}, vol. 3, no. 1-2, pp. 1091--1102, 2010.

\bibitem{ref_On-the-spatiotemporal-burstiness}
T. Lappas, M. R. Vieira, D. Gunopulos, and V. J. Tsotras. ``On the spatiotemporal burstiness of terms,'' \textit{PVLDB}, vol. 5, no. 9, pp. 836--847, 2012.

%%=================driving=======================
\bibitem{ref_Real-time-obstacles-detection}
W. Song, Y. Yang, M. Fu, F. Qiu, and M. Wang,
``Real-time obstacles detection and status classification for collision warning
in a vehicle active safety system,''
\textit{IEEE Transactions on Intelligent Transportation Systems}, vol. PP, no. 99, pp. 1--16,
May. 2017.

\bibitem{ref_3D-Geometry-from-Planar}
H. S. Sawhney, ``3D geometry from planar parallax,'' in \textit{Proc. CVPR}, 1994.

\bibitem{ref_Real-Time-Obstacle-Detection}
R. Labayrade, D. Aubert, and J. P. Tarel, ``Real time obstacle detection in stereovision
on non flat road geometry through "vdisparity" representation,''
in \textit{Proc. IV Symposium}, 2002.

\bibitem{ref_Obstacle-Detection-Using-Sparse}
S. Kramm and A. Bensrhair, ``Obstacle detection using sparse stereovision and clustering techniques,''
in \textit{Proc. IV Symposium}, 2012.

\bibitem{ref_Obstacle-detection-based-on}
Z. Zhang, R. Weiss, and A. Hanson, ``Obstacle detection based on qualitative
and quantitative 3d reconstruction,''
\textit{TPAMI}, vol. 19, no. 1, pp. 15--26, 1997.

\bibitem{ref_Lost-and-found}
P. Pinggera, S. Ramos, S. Gehrig, U. Franke, C. Rother, and R. Mester,
``Lost and found: Detecting small road hazards for self-driving vehicles,''
in \textit{Proc. 2016 IEEE/RSJ International Conference on Intelligent Robots and Systems (IROS)},
Daejeon, Korea, Oct. 2016, pp.1099--1106.


\bibitem{ref_A-real-time-low}
S. K. Gehrig, F. Eberli, and T. Meyer, ``A real-time low-power stereo vision engine using semi-global matching,''
in \textit{Proc. 7th Int. Conf. Comput. Vis. Syst.}, vol. 5815, 2009, pp. 134--143.

\bibitem{ref_A-head-wearable-short}
H. Badino and T. Kanade, ``A head-wearable short-baseline stereo system for the
simultaneous estimation of structure and motion,''
in \textit{Proc. MVA}, 2011, pp. 185--189.

\bibitem{ref_Model-based-vehicle-detection}
P. Anna, and S. Thrun. ``Model based vehicle detection and tracking for autonomous urban driving,''
\textit{Auton. Robot.}, vol. 26, no. 2-3, pp. 123--139, 2009.

\bibitem{ref_Object-detection-and-tracking}
Y. Ye, L. Fu, and B. Li,
''Object detection and tracking using multi-layer laser for autonomous urban driving,''
in \textit{Proc. 2016 IEEE 19th International Conference on Intelligent Transportation Systems (ITSC)},
Rio de Janeiro, Brazil, Nov. 2016, pp. 259--264.

\bibitem{ref_RefineNet}
R. N. Rajaram, E. O. Bar, and M. M. Trivedi,
``RefineNet: Refining object detectors for autonomous driving,''
\textit{IEEE Trans. Intell. Veh.},
vol. 1, no. 4, pp. 358--368, Dec. 2016.

\bibitem{ref_Selective-attention-for-detection}
A. Unterholzner and H. Wuensche,
``Selective attention for detection and tracking of road-networks in autonomous driving,''
in \textit{Proc. 2013 IEEE Intelligent Vehicles Symposium (IV)},  Gold Coast, Australia,
June. 2013, pp. 277--284.

\bibitem{ref_Simultaneous-localization-and-mapbuilding}
A. J. Davison and D. W. Murray, ``Simultaneous localization and mapbuilding using active vision,''
\textit{ IEEE Trans. Pattern Anal. Mach. Intell.}, vol. 24, pp. 865--880, 2002.

\bibitem{ref_Image-based-automatic-road}
C. Premachandra, H. Waruna, H. Premachandra, and C. D. Parape,
``Image based automatic road surface crack detection for
achieving smooth driving on deformed roads,'' in
\textit{ Proc. 2013 IEEE International Conference on Systems, Man, and Cybernetics(SMC)},
Manchester, United Kingdom, Oct. 2013, pp. 4018--4023.

\bibitem{ref_Automatic-crack-detection-on}
H. Oliveira and P. L. Correia, ``Automatic crack detection on road imagery
 using anisotropic diffusion and regional linkage,'' in
\textit{ Proc. 18th European Signal Processing Conference}, 2010, pp. 274--278.

\bibitem{ref_Virtuous-visionbased-road-transportation}
M. A. Sotelo, F. J. Rodriguez, and L. Magdalena, ``Virtuous: visionbased road
transportation for unmanned operation on urban-like scenarios,''
\textit{IEEE Trans. Intell. Transp. Syst.},
vol. 5, no. 2, pp. 69--83, June. 2004.

\bibitem{ref_A-color-vision-based-lane}
M. A. Sotelo, F. J. Rodriguez, L. Magdalena, L. M. Bergasa, and L. Boquete, ``A color vision-based
lane tracking system for autonomous driving on unmarked roads,''
\textit{Auton. Robot.}, vol. 16, no. 1, pp. 95--116, 2004.

\bibitem{ref_Road-curb-and-lanes-detection}
C. Fernandez, R. Izquierdo, D. F. Llorca, M. A. Sotelo,
``Road curb and lanes detection for autonomous driving on urban scenarios,'' in
\textit{ Proc. 2014 IEEE 17th International Conference on Intelligent Transportation Systems (ITSC)},
Qingdao, China, Oct. 2014, pp. 1964--1969.

%%%=======================
%\vspace{-1mm}
%\begin{shaded}
%\vspace{-3mm}
\bibitem{ref_Learning-Probabilistic}
D. Campo, M. Baydoun, P. Marin, D. Martin, L. Marcenaro, A. Escalera, and C. Regazzoni,
``Learning probabilistic awareness models for detecting abnormalities in vehicle motions,''
\textit{IEEE Trans. Intell. Transp. Syst.},
pp. 1--13, Apr. 2019.

\bibitem{ref_A-PSO-and-BFO-based}
G. Wang, J. Guo, Y. Chen, Y. Li, and Q. Xu,
``A PSO and BFO-based learning strategy applied to faster R-CNN for object detection in autonomous driving,''
\textit{IEEE Access}, vol. 7, pp. 18840--18859, Feb. 2019.

\bibitem{ref_3D-Object-Proposals}
X. Chen, K. Kundu, Y. Zhu, H. Ma, S. Fidler, and R. Urtasun,
``3D Object Proposals Using Stereo Imagery for Accurate Object Class Detection,''
\textit{IEEE Trans. Pattern Anal. Mach. Intell.},
vol. 40, no. 5, pp. 1259--1272, May. 2018.
%\end{shaded}


%%=====IoT==================================
\bibitem{ref_Recursive-principal-component-analysis}
T. Yu, X. Wang, and A. Shami,
``Recursive principal component analysis based data outlier
detection and sensor data aggregation in iot systems,'' \textit{IEEE Internet Things J.},
vol. PP, no. 99, pp. 1--10, 2017.

\bibitem{ref_Outlier-Detection-Techniques-for}
Y. Zhang, N. Meratnia, and P. Havinga, ``Outlier detection techniques for wireless
sensor networks: a survey,''
\textit{IEEE Commun. Surveys Tuts.}, vol. 12, no. 2, pp. 159--170, 2010.

\bibitem{ref_Research-Directions-for-the}
J. A. Stankovic, ``Research Directions for the Internet of Things,''
\textit{IEEE Internet Things J.}, vol. 1, no. 1, pp. 3--9, 2014.


\bibitem{ref_A-lightweight-anomaly-mining}
Y. Liu and Q. Wu,
``A lightweight anomaly mining algorithm in the Internet of Things,'' in
\textit{Proc. 2014 IEEE 5th International Conference on Software Engineering and Service Science},
Beijing, China, June. 2014, pp. 1142--1145.

\bibitem{ref_Detection-of-anomalies-in}
V. A. Desnitsky, I. V. Kotenko, and S. B. Nogin,
``Detection of anomalies in data for monitoring of security components in the internet of things,'' in
\textit{Proc. 2015 XVIII International Conference on Soft Computing and Measurements (SCM)},
St. Petersburg, Russia, May. 2015,
pp. 189--192.

\bibitem{ref_Anomaly-detection-and-monitoring}
D. Stiawan, M. Y. Idris, R. F. Malik, S. Nurmaini, and R. Budiarto,
``Anomaly detection and monitoring in internet of things communication,''
in \textit{Proc. 2016 8th International Conference on Information Technology and Electrical Engineering (ICITEE)},
Yogyakarta, Indonesia, Oct. 2016,
pp. 1--4.

\bibitem{ref_Clustering-for-road-damage}
J. Takahashi, D. Shioiri, Y. Shida, Y. Kobana, R. Suzuki, Y. Kobayashi, N. Isoyama, G. Lopez, and Y. Tobe,
``Clustering for road damage locations obtained by smartphone accelerometers,''
in \textit{Proc. the Second International Conference on IoT in Urban Space(UrbIoT '16)},
Tokyo, Japan, May, 2016, pp. 89--91.

\bibitem{ref_Urban-anomaly-detection}
J. Borges, T.Riedel, and M. Beigl,
``Urban anomaly detection: A use-case for participatory infra-structure monitoring,''
in \textit{Proc. The Second International Conference on IoT in Urban Space(UrbIoT '16)},
Tokyo, Japan, May, 2016, pp. 36--38.

\bibitem{ref_Urban-sensing-and-smart}
M. Shahriar and M. Rahman,
``Urban sensing and smart home energy optimisations: A machine learning approach,''
\textit{Proc. 2015 International Workshop on Internet of Things Towards Applications (IoT--App '15)},
Seoul, South Korea, pp. 19--22.

\bibitem{ref_Performance-analysis-of-anomaly}
N. Rakesh,
``Performance analysis of anomaly detection of different IoT datasets using cloud micro services,''
in \textit{Proc. 2016 International Conference on Inventive Computation Technologies (ICICT)},
Coimbatore, India, Aug. 2016,
pp. 1--5.

\bibitem{ref_Towards-an-emulated-IoT}
S. Brady, A. Hava, P. Perry, J. Murphy, D. Magoni, and A. Dominguez,
``Towards an emulated IoT test environment for anomaly detection using NEMU,''
in \textit{Proc. 2017 Global Internet of Things Summit (GIoTS)},
Geneva, Switzerland, June. 2017,
pp. 1--6.

%%==========
%\vspace{-1mm}
%\begin{shaded}
%\vspace{-3mm}
\bibitem{ref_Anomaly-Detection-and-Attribution}
I. Nevat, D. M. Divakaran, S. G. Nagarajan, P. Zhang, L. Su, L. L. Ko, and V. L. L. Thing,
``Anomaly detection and attribution in networks with temporally correlated traffic,''
\textit{IEEE/ACM Trans. Netw.},
vol. 26, no. 1, pp. 131--144, Feb. 2018.

\bibitem{ref_Outlier-Dirichlet-Mixture-Mechanism}
N. Moustafa, K. R. Choo, I. Radwan, S. Camtepe,
``Outlier Dirichlet mixture mechanism: Adversarial statistical learning for anomaly detection in the fog,''
\textit{IEEE Trans. Inf. Forensics Security},
vol. 14, no. 8, pp. 1975--1987, Aug. 2019.

\bibitem{ref_An-Intelligent-Outlier-Detection-Method-With}
X. Deng, P. Jiang, X. Peng, and C. Mi,
``An intelligent outlier detection method with one class support tucker machine and genetic algorithm toward big sensor data in Internet of Things,''
\textit{IEEE Trans. Ind. Electron.},
vol. 66, no. 6, pp. 4672--4683, Jun. 2019.
%\end{shaded}

\end{thebibliography}
\end{document}